\documentclass[aps,prxquantum,twocolumn,superscriptaddress,floatfix,nofootinbib,longbibliography]{revtex4-2}
\usepackage{amsmath,amssymb,amsthm,graphicx,bm,dsfont}
\usepackage{tikz}
\usetikzlibrary{arrows.meta,shapes.geometric,decorations.pathreplacing}
\usepackage{quantikz}
\usepackage[colorlinks=true,linkcolor=blue,citecolor=blue]{hyperref}
\usepackage{orcidlink}
\newcommand{\Tr}{\mathrm{Tr}}
\newcommand{\id}{\mathds{1}}
\newtheorem{theorem}{Theorem}
\newtheorem{corollary}{Corollary}
\newtheorem{proposition}{Proposition}

\begin{document}

\title{Quantized memory of measurement records near the light cone of quantum circuits}

\author{Jamal Slim\orcidlink{0000-0002-9418-8459}}
\email{jamal.slim@desy.de}
\affiliation{Deutsches Elektronen-Synchrotron DESY, 22603 Hamburg, Germany}

\date{\today}

\begin{abstract}
Quantum processors now measure qubits in the middle of a circuit, and the
stream of outcomes is the raw output of error correction, measurement-based
state preparation and monitored dynamics. For a
scrambling circuit, predicting the statistics of this stream is as hard as
simulating the processor, so there is no exact reference to check it against
once the device outgrows classical simulation. We show that near the light
cone the problem becomes exactly solvable, for any two-qubit gate. In a brickwork circuit with one polarized qubit in a maximally mixed
background, measurements along the edge of the light cone form a
two-state Markov chain, and measurements $d$ sites inside the edge are
generated by $\lfloor d/2\rfloor+1$ qubits that move with it, while the rest
of the chain, however large, acts on them as a memoryless bath. The memory of
the record is therefore quantized, one bit on the edge and at most one more
for every site of displacement, a bound that generic gates attain. The same
causal structure determines which hardware noise the record can see, removes
that sensitivity at no gate cost, and yields a tomography-free witness of
dual-unitarity. On a 156-qubit superconducting processor, a demonstration
with its analysis specified in advance finds the predicted memory one and two
sites inside the light cone, and none where the theory forbids it. This gives
dynamic circuits an exact reference at any system size, and shows that the
collision models of open-system theory arise exactly inside chaotic
dynamics.
\end{abstract}

\maketitle

\section{Introduction}

A quantum processor that measures some of its qubits in the middle of a
circuit produces a stream of classical bits. That stream is the raw output of
quantum error correction, where syndromes are measured round after round, of
measurement-based preparation of long-range entangled states, and of the
monitored dynamics in which measurements compete with scrambling. In each
case the record is what the device reports, and whether the device works is
a question about the statistics of that record. For a scrambling circuit the
question is hard to answer. Predicting the joint distribution of a sequence
of mid-circuit outcomes is in general as hard as simulating the circuit that
produced them, so the reference against which a record could be checked
disappears precisely where it is most needed, once the device runs circuits
beyond the reach of classical simulation~\cite{HangleiterEisert}. This paper
asks whether any part of that record can be predicted exactly, for an
arbitrary gate and at any system size, and answers that near the light cone
of the circuit all of it can.

The tools that exist address the problem only in part. Randomized and cycle
benchmarking characterize gates, and dedicated protocols now measure the
errors of mid-circuit measurements and the crosstalk they cause on spectator
qubits~\cite{Govia,Hothem,HinesProctor,ZhangCB,ProctorReview}, but none of
them provides an exact reference for the dynamics of a scrambling,
non-Clifford circuit as a whole. Exactly solvable circuits do, but only once
the gate has been fine tuned. Dual-unitary circuits admit closed-form
correlation functions at any depth~\cite{BKP2019,Piroli2020,RMP} and have
served as a benchmark for a 91-qubit processor~\cite{Anchor}. Their
generalizations are organized as hierarchies of conditions on the
gate~\cite{YuWangKos,Rampp2026}, exact hidden-Markovian subsystem dynamics
has been engineered for gates that satisfy a solvability
condition~\cite{WangPRL}, and the recently introduced influence-solvable
circuits contain most of these families as special cases~\cite{HubnerKim}.
Some tractability, however, owes nothing to the gate. In a brickwork circuit
at infinite temperature, the reduced state of a few sites that ride the light
cone evolves under a channel whose dimension does not depend on the length of
the chain, whatever the gates are~\cite{BKP2019,ClaeysLamacraft,SinghDooley},
and two-point functions on the edge of the cone follow from unitarity alone.
Measurements have entered this picture only in special settings. In
monitored circuits of XYZ gates, post-selected light-cone correlations can be
obtained from typical correlations of a different unitary
circuit~\cite{PostSel}, and dual-unitary circuits measured in a unitary error
basis remain solvable, although every outcome is then equally likely and the
record carries no information~\cite{CHVL}. The object a device actually
produces, the joint distribution of a sequence of mid-circuit outcomes with
the back-action of every earlier measurement included, had no exact
description for an arbitrary gate.

Our answer rests on a change of frame. Viewed along its diagonals, a
brickwork circuit is a cascade of left-moving lines in which information
passes from each line only to the line behind it. An observer who reads the
chain $d$ sites inside the edge of the light cone, moving with it, interacts
with $\lfloor d/2\rfloor+1$ of these lines and with nothing else, and for
this observer the remaining qubits of a scrambling chain form an exactly
memoryless bath (Theorem~\ref{thm:comoving}). On the edge itself the record
is a two-state Markov chain with $4\times4$ transfer matrices, for any gate
and any system size (Theorem~\ref{thm:edge} and
Corollary~\ref{cor:chain}). Protection against noise follows the same causal
structure. Channels acting behind the measured ray are invisible, channels
acting ahead of it are invisible only if they are unital, and a Pauli twirl
applied to each site just before it joins the ray restores the exact law at
no gate cost (Theorem~\ref{thm:noise}). The frame also fixes how much the
record can remember. Its memory, measured by the rank of its joint
distribution across a cut in time, is at most $d+1$ bits
(Corollary~\ref{cor:ladder}), and generic gates reach the cap, as we prove
for displacements of up to six sites (Proposition~\ref{prop:generic}). One
site inside the edge the measured qubit is a passing ancilla, the co-moving
qubit is never reset, and the record is not Markovian at any finite order,
even though two bits generate it exactly. This is the structure of a
collision model~\cite{Ciccarello} or of a cascaded quantum
system~\cite{Gardiner,Carmichael}, arising here as an exact property of
many-body dynamics rather than as a modelling assumption. Readouts slower
than the edge fall behind through more and more lines, and their memory
grows with the length of the record. Dual-unitary gates are exceptional.
Their records carry at most one bit at every displacement we examined and
none at all one site inside the edge, which turns that record into a witness
of dual-unitarity that single-qubit noise cannot fake.

Three consequences reach beyond the model. For the validation of quantum
hardware, the record law is an exact reference for a scrambling circuit with
mid-circuit measurements whose cost does not grow with the number of qubits,
in the regime of dynamic circuits in which such references have been
missing, together with a precise account of the noise it can see and a
tomography-free test of the executed gate. For the theory of many-body
dynamics, it identifies an exact structure in chaotic circuits that does not
depend on the gate, and it turns the question of where the output of
monitored dynamics becomes hard to predict into a count of light-cone lines.
For open-system theory, it shows that the collision models used to describe
system-bath dynamics arise exactly inside a generic scrambling chain, with the
bath's memory quantized by geometry.

We test these predictions on the superconducting processor
\textit{ibm\_fez}, accessed through the cloud, in three steps. Twelve
records taken before the present analysis obey the Markov structure once the
transition kernel is allowed to vary from step to step, as it must on a
device in which every step uses a different pair of physical qubits. A
single interleaved demonstration, analysed against acceptance criteria set
in advance, then shows that relaxation injected ahead of the ray corrupts the
record only in the absence of the late twirl, and that the witness places the
executed dual-unitary gate within $0.011$\,rad of the dual-unitary manifold.
Finally, the records of that demonstration revealed the memory predicted one
site inside the edge, and a second demonstration, whose analysis was specified
before its data existed, confirms it for two scrambling gates one and two sites inside the
edge, together with its absence wherever the theory predicts none.
Repetitions of the registered demonstration on two further processors find
the memory wherever it is predicted and its absence wherever it is not.

The records studied here are classical in a precise sense, because the
record law is its own sampler. They belong to the effectively classical
temporal correlations identified in the influence-matrix
literature~\cite{LSA,WVA,VA}, and in the language of process tensors they have
a memory of fixed size~\cite{Pollock}. Dual-unitary circuits are known to act
as perfectly Markovian baths for a subsystem at rest~\cite{LSA}, and temporal
entanglement along tilted paths has been analysed for solvable and chaotic
circuits after quenches from pure states~\cite{FZB}. Our statement is
complementary. At infinite temperature and for an arbitrary gate, the bath
seen from a frame that moves with the edge is memoryless, and frames differ
only in how many qubits the observer has to carry along. The value of the
construction lies in an exact reference that does not degrade with
system size, not in any computational advantage.

The paper is organized as follows. Section~\ref{sec:law} derives the record
law on the edge, Sec.~\ref{sec:noise} gives the causal account of noise, and
Sec.~\ref{sec:comoving} develops the co-moving frame, the memory of the
record and the dual-unitarity witness. Section~\ref{sec:verify} summarizes
the checks against exact simulation, Sec.~\ref{sec:hardware} reports the
demonstrations on hardware, and Sec.~\ref{sec:discussion} discusses scope
and open questions. Proofs, the documentation of the cloud platform and the
details of the demonstrations are collected in the appendices.

\section{Measurement records on the light-cone edge}
\label{sec:law}

\begin{figure}[t]\centering
\input{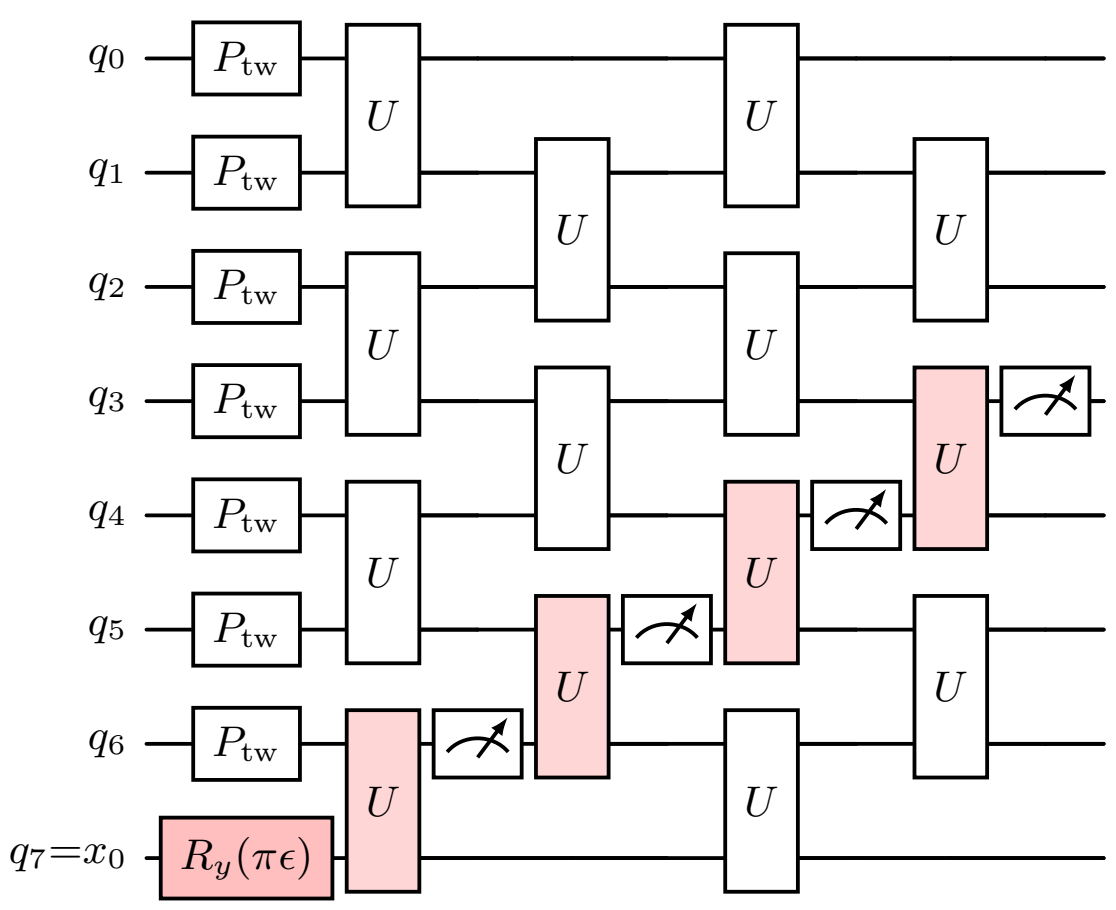}
\caption{Measurement circuit for $n=8$, $S=4$ and $x_0=7$. The germ site
$x_0$ is prepared by a rotation $R_y(\pi\epsilon)$, and every other site is
randomized from shot to shot ($P_{\rm tw}$), which realizes the maximally
mixed background. Gates act in brickwork layers, and after half-step $h$ the
site $x_0-h$ is measured in $Z$. Shaded gates lie on the edge of the light
cone. As drawn, every site is randomized at $t=0$, as in the recorded rails of
Sec.~\ref{sec:rails}. In the late-twirl protocol of
Theorem~\ref{thm:noise}(iii), each rail site $x_0-h$ is instead randomized
immediately before its shaded gate.}
\label{fig:circuit}
\end{figure}

Consider a chain of $n$ qubits prepared in the state
\begin{equation*}
\rho_0=\tfrac12(\id+\bm r\cdot\bm\sigma)_{x_0}\otimes
\bigotimes_{q\neq x_0}\tfrac{\id}{2},
\end{equation*}
in which a single site $x_0$, which we call the germ, carries the Bloch
vector $\bm r$ and every other site is maximally mixed. The chain evolves
under a brickwork of two-qubit gates, with even bonds acting on odd
half-steps and odd bonds on even half-steps. After half-step $h=1,\dots,S$
we measure site $x_0-h$ in the $Z$ basis and obtain a record
$\bm m=(m_1,\dots,m_S)$ with $m_h=\pm1$. The measured sites trace the left
edge of the light cone of the germ, one site per half-step, which is the
largest speed the circuit allows. We call this sequence of sites a rail
(Fig.~\ref{fig:circuit}).

Two $4\times4$ matrices describe the action of a single gate on the edge. For
a gate $U$ on the sites $(a,a+1)$, with $a$ the first tensor factor, the
Heisenberg-picture light-cone channel has the real matrix elements
\begin{equation}
\mathcal M_{ba}=\tfrac12\Tr\!\big[\sigma_b\,\Tr_1[U^\dagger(\sigma_a\otimes
\id)U]\big],\qquad a,b\in\{\id,X,Y,Z\}.
\label{eq:chan}
\end{equation}
The Schr\"odinger-picture edge propagator, which carries the state of the
right site over to the left site while the left site enters maximally mixed,
$\rho\mapsto\Tr_2[U(\tfrac{\id}{2}\otimes\rho)U^\dagger]$, has the Bloch matrix
$A=\mathcal M^{\mathsf T}$ (Appendix~\ref{app:transpose}). The transpose is
easy to overlook and matters in practice. A germ polarized along $z$ cannot
tell $A$ from $\mathcal M$, so a numerical check of the law needs germs with
transverse components (Table~\ref{tab:ver}).

\begin{theorem}[Edge record]
\label{thm:edge}
Suppose the germ sits on the sublattice paired with the first brickwork
layer, so that $x_0$ and $x_0-1$ share a gate at $h=1$. Then the record
distribution is
\begin{equation}
p(m_1\ldots m_S)= e_{\id}^{\mathsf T}\,K_{m_S}A_S\cdots K_{m_1}A_1\,v_0,
\qquad v_0=(1,\bm r),
\label{eq:law}
\end{equation}
where $A_h$ is built from the gate on the edge at half-step $h$,
$e_{\id}^{\mathsf T}$ selects the identity component, and each measurement
updates the ray by
\begin{equation}
K_m=\frac12\begin{pmatrix}1&0&0&m\\0&0&0&0\\0&0&0&0\\m&0&0&1\end{pmatrix}.
\label{eq:K}
\end{equation}
\end{theorem}

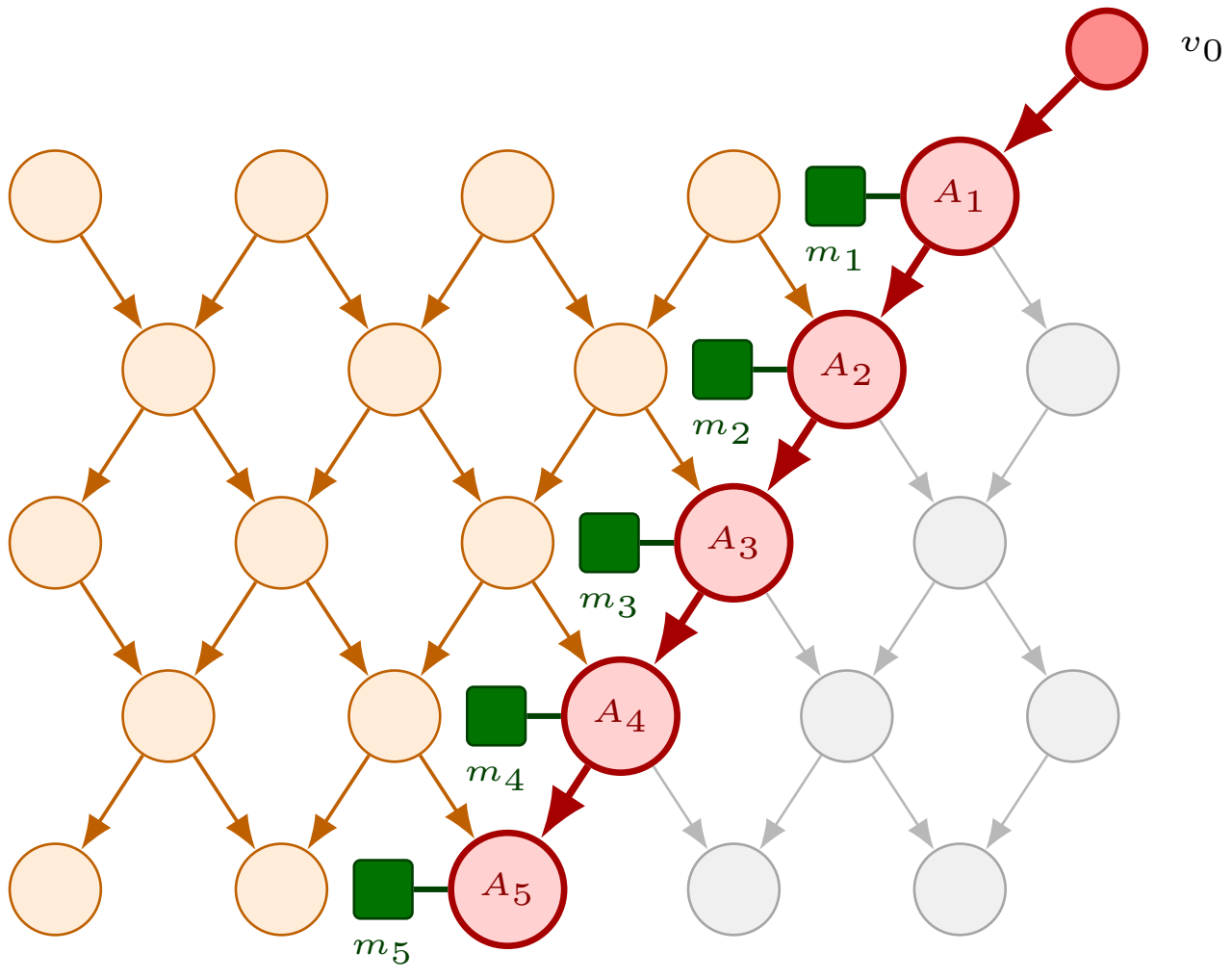
\begin{figure}[t]\centering
\resizebox{\columnwidth}{!}{% Causal graph of a brickwork circuit, 5 half-steps.
% Position index p in units of one site: odd layers p = 0,2,...,8, even layers
% p = 1,3,...,9.  Each gate feeds exactly the two gates at p-1 and p+1 of the
% next layer.  The ray gate of half-step h sits at p = 9-h, so the ray moves
% one site per half-step.  p < 9-h is ahead of the ray, p > 9-h behind it.
\begin{tikzpicture}[
  x=0.62cm, y=0.95cm, >=Latex, line join=round,
  bnode/.style={circle, draw=black!35, fill=black!6, inner sep=0pt, minimum size=5.0mm},
  anode/.style={circle, draw=orange!75!black, fill=orange!15, inner sep=0pt, minimum size=5.0mm},
  rnode/.style={circle, draw=red!65!black, fill=red!18, line width=1.0pt, inner sep=0pt, minimum size=6.2mm},
  mnode/.style={rectangle, rounded corners=1pt, draw=green!25!black, fill=green!45!black, inner sep=0pt, minimum size=3.2mm},
  dead/.style={black!28, line width=0.4pt},
  feed/.style={orange!75!black, line width=0.55pt},
  live/.style={red!65!black, line width=1.15pt},
  tiny/.style={font=\tiny},
]
\foreach \h in {1,...,5}{
  \foreach \k in {0,...,4}{
    \pgfmathtruncatemacro{\p}{2*\k+mod(\h+1,2)}
    \pgfmathtruncatemacro{\r}{9-\h}
    \ifnum\p=\r
      \node[rnode] (n\h-\p) at (\p,-\h) {};
      \node[tiny,red!55!black] at (\p,-\h) {$A_{\h}$};
    \else\ifnum\p<\r
      \node[anode] (n\h-\p) at (\p,-\h) {};
    \else
      \node[bnode] (n\h-\p) at (\p,-\h) {};
    \fi\fi
  }
}
\foreach \h in {1,...,4}{
  \pgfmathtruncatemacro{\hn}{\h+1}
  \pgfmathtruncatemacro{\r}{9-\h}
  \pgfmathtruncatemacro{\rn}{9-\hn}
  \foreach \k in {0,...,4}{
    \pgfmathtruncatemacro{\p}{2*\k+mod(\h+1,2)}
    \foreach \s in {-1,1}{
      \pgfmathtruncatemacro{\q}{\p+\s}
      \pgfmathtruncatemacro{\ok}{(\q>=0 && \q<=9) ? 1 : 0}
      \ifnum\ok=1
        \ifnum\p=\r
          \ifnum\q=\rn \draw[live,->] (n\h-\p) -- (n\hn-\q);
          \else \draw[dead,->] (n\h-\p) -- (n\hn-\q); \fi
        \else\ifnum\p<\r
          \draw[feed,->] (n\h-\p) -- (n\hn-\q);
        \else
          \draw[dead,->] (n\h-\p) -- (n\hn-\q);
        \fi\fi
      \fi
    }
  }
}
\node[rnode,fill=red!45,minimum size=4.2mm] (germ) at (9.3,-0.15) {};
\node[tiny,anchor=west] at (9.75,-0.15) {$v_0$};
\draw[live,->] (germ) -- (n1-8);
\foreach \h in {1,...,5}{
  \pgfmathtruncatemacro{\r}{9-\h}
  \node[mnode] (m\h) at (\r-1.1,-\h) {};
  \node[tiny,green!25!black,anchor=north] at (\r-1.1,-\h-0.17) {$m_{\h}$};
  \draw[green!25!black,line width=0.9pt] (n\h-\r) -- (m\h);
}
\end{tikzpicture}}
\caption{Causal structure of the circuit, with gates as nodes and light-cone
dependence as arrows. The germ $v_0$ reaches each record bit $m_h$ along a
single path, through the edge gates $A_h$ (red). Grey nodes lie behind the
ray and have no directed path to any $m_h$, so no channel acting there can
change the record [Theorem~\ref{thm:noise}(i)]. Orange nodes lie ahead of
the ray and reach the record only through the incoming leg of an edge gate,
which arrives maximally mixed. Unitary gates and unital channels there leave
the record unchanged, whereas non-unital channels do not
[Theorem~\ref{thm:noise}(ii)]. The graph makes no reference to the content of
the gates, which is why Theorem~\ref{thm:edge} holds for any unitary.}
\label{fig:dag}
\end{figure}

Nothing in Eq.~\eqref{eq:law} refers to the gates beyond their unitarity,
and nothing refers to the length of the chain. A likelihood costs $O(S)$
products of $4\times4$ matrices, and the gates may differ from one half-step
to the next. The reason is causal (Fig.~\ref{fig:dag}). A brickwork layer
moves information by at most one site, so the edge is the only path of
maximal speed from the germ to each measurement, and every other path arrives
too late. The back-action of a measurement obeys the same bound. As a result,
the site that enters each edge gate from the left lies strictly outside the
forward light cone of the germ and of all earlier measurements, and it
arrives maximally mixed and uncorrelated with the ray
(Appendix~\ref{app:recordlaw}). In this setting the maximally mixed
background does the work that dual-unitarity does elsewhere. The parity
condition is essential. With the germ on the other sublattice, the germ and
the first measured site belong to different gates, and Eq.~\eqref{eq:law}
fails at the percent level (Table~\ref{tab:ver}).

For expectation values this is standard. Unitarity contracts every leg
outside the backward light cone, so two-point functions on the edge are
products of $\mathcal M$ for any brickwork circuit~\cite{BKP2019,SinghDooley}.
A record is a different object. A projective measurement is not a
contraction, $K_m$ is not unital, and every outcome carries the back-action
of the measurements that preceded it. What Theorem~\ref{thm:edge} adds is
that the joint distribution of all outcomes nevertheless remains an ordered
product of $4\times4$ matrices, for any gate.

The structure simplifies further after the first measurement. The update
$K_m$ projects the ray onto $|m\rangle$ and removes its $X$ and $Y$
components, so only the first outcome sees the transverse part of the germ,
and from then on the hidden state is simply the last outcome
(Appendix~\ref{app:cor1}).

\begin{corollary}[Monitored chain]
\label{cor:chain}
Under the hypotheses of Theorem~\ref{thm:edge},
\begin{equation}
p(\bm m)=\tfrac12\big(1+m_1\mu_1\big)\prod_{h=2}^{S}
\tfrac12\big[1+m_h\,(t_h+s_h\,m_{h-1})\big],
\label{eq:chain}
\end{equation}
with $\mu_1=[A_1v_0]_Z$, $t_h=[A_h]_{Z\id}$ and $s_h=[A_h]_{ZZ}$.
\end{corollary}

For unitary gates the bias $t_h$ vanishes and $s_h$ is the monitored decay
rate, so that $\langle m_jm_k\rangle=\prod_{h=j+1}^{k}s_h$. For the
dual-unitary gate used in our demonstrations $s=0.607$, and the light-cone
correlations at lags one to three are $0.607$, $0.368$ and $0.224$. The
gate-invariant content of $\mathcal M$ lies in its singular values,
$\{1,1,\sin2J,\sin2J\}$ with $\sin2J=0.731$ for this gate, whereas its
eigenvalues, $\{1,0.734,-0.561\pm0.643i\}$, change under local unitaries and
carry no invariant meaning. A single-qubit channel acting on the rail changes
only $(\mu_1,t_h,s_h)$, and a readout assignment error turns the chain into a
hidden two-state chain. Equation~\eqref{eq:chain} with free, step-dependent
parameters and assignment errors is therefore the most general record that
the causal structure allows, and it is the null hypothesis we test on
hardware in Sec.~\ref{sec:rails}.

\section{Noise and causality}
\label{sec:noise}

Real devices add noise to every gate, idle period and readout, and an exact
law is only useful if one knows which of these errors it is blind to. The
light-cone structure answers this question sharply.

\begin{theorem}[Causal dichotomy and late twirl]
\label{thm:noise}
Let the ray occupy site $x_0-h$ after half-step $h$.
(i)~Arbitrary completely positive trace-preserving maps, unital or not,
acting on sites $q>x_0-h$ behind the ray leave the record distribution
exactly invariant.
(ii)~Maps acting on sites $q<x_0-h$, which have yet to join the rail, leave
it invariant if and only if they are unital.
(iii)~If each incoming site $x_0-h$ receives a uniformly random Pauli
operator immediately before its edge gate, the noisy record law
\begin{equation}
p(\bm m)=e_{\id}^{\mathsf T} K^{(q)}_{m_S}\mathcal N_S A_S\cdots
K^{(q)}_{m_1}\mathcal N_1 A_1 v_0
\label{eq:noisylaw}
\end{equation}
holds exactly, whatever happened to the incoming sites before they joined the
ray. Here $\mathcal N_h$ is an arbitrary single-qubit channel on the rail
site and $K^{(q)}_m=(1-q)K_m+qK_{-m}$ accounts for a readout assignment
error $q$.
\end{theorem}

\begin{figure}[t]
\includegraphics[width=\columnwidth]{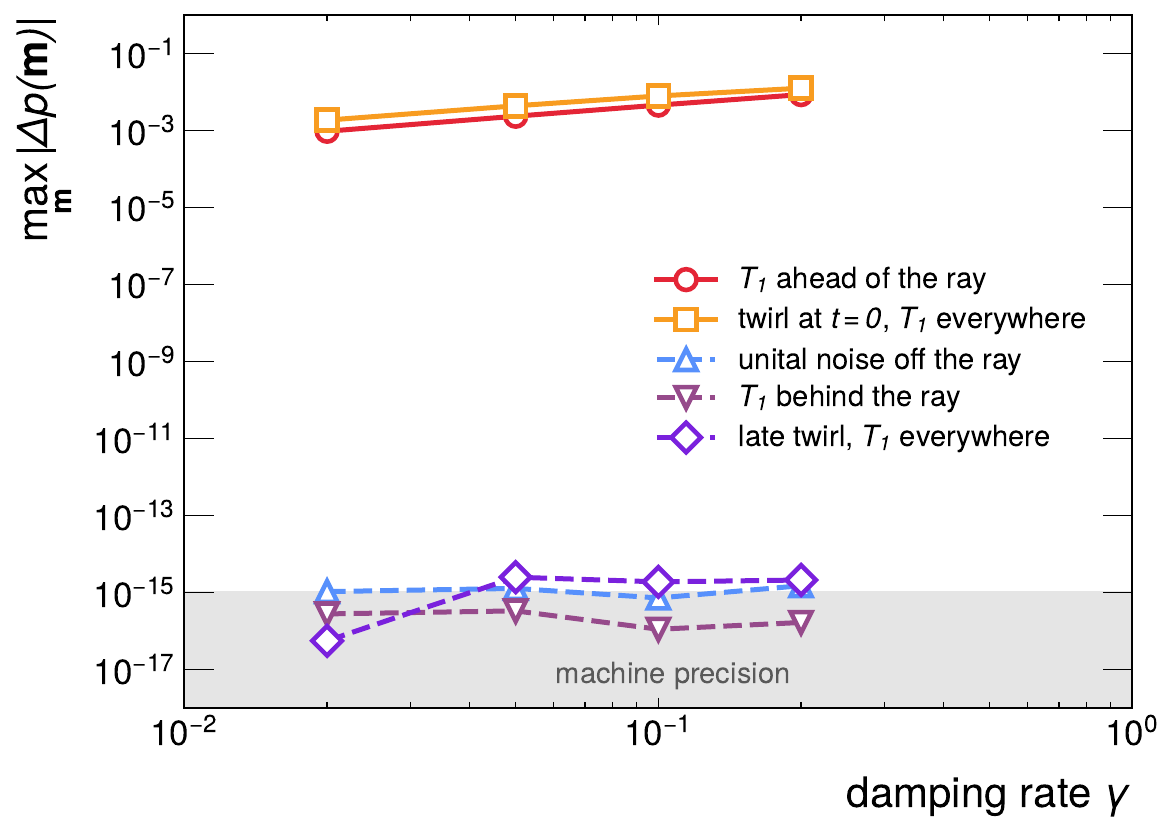}
\caption{The causal dichotomy of Theorem~\ref{thm:noise} in full
density-matrix simulation. Deviation of the record distribution from
Eq.~\eqref{eq:law} as a function of the damping rate $\gamma$, for $T_1$
damping behind the ray, unital noise off the ray, $T_1$ damping ahead of the
ray, and the two twirl schedules. Only non-unital noise ahead of the ray is
visible, and the late twirl removes it exactly.}
\label{fig:dich}
\end{figure}

Parts (i) and (ii) can be read off Fig.~\ref{fig:dag}. A site behind the ray
never re-enters the backward light cone of a later measurement, so nothing
done to it can matter. A site ahead of the ray reaches the record only as the
incoming leg of an edge gate, and Theorem~\ref{thm:edge} needs that leg to
arrive maximally mixed, which a channel guarantees exactly when it is unital
(Appendix~\ref{app:dichotomy}). Part (ii) is the one real vulnerability of
the protocol, and it is not an academic one. On hardware the qubits that will
later join the rail idle through every earlier measurement window, and
amplitude damping during these idle periods is non-unital. In a full
density-matrix simulation with $T_1$ damping on every site, the record
deviates from Eq.~\eqref{eq:law} at first order in the damping rate, and no
law of the form of Eq.~\eqref{eq:law} reproduces the deviation
(Fig.~\ref{fig:dich}).

Part (iii) removes this vulnerability. Averaging over the four Pauli
operators on a site maps that site, together with every correlation it has
with the rest of the chain, to $\tfrac{\id}{2}\otimes\Tr_q\rho$, which is
precisely the hypothesis of Theorem~\ref{thm:edge}. Applied just before the
edge gate instead of at the start of the circuit, the twirl erases whatever
the incoming site picked up while it waited. It costs no two-qubit gate and
at most one single-qubit gate per rail site. It also makes the preparation of
the rail sites irrelevant, so they may be left in $|0\rangle$, the most
strongly polarized state there is.

Dual-unitary gates are nearly immune to the exposure even without the late
twirl. Their stay channel $\rho\mapsto\Tr_2[U(\rho\otimes\tfrac{\id}{2})
U^\dagger]$ is completely depolarizing, so a polarized incoming leg affects
the record only through terms bilinear in its own polarization and that of
the ray. In simulation, relaxation of strength $\gamma=0.2$ on every site
shifts the transition kernel of the dual-unitary gate by at most $0.011$, and
that of a Haar-random gate by up to $0.042$. For this reason the hardware
test of the late twirl in Sec.~\ref{sec:idle} uses a Haar-random gate.

\section{The co-moving frame and the memory of the record}
\label{sec:comoving}

Now move the readout $d$ sites into the cone, so that after half-step $h$ it
measures site $x_0-h+d$. The germ then reaches each measurement along many
paths, and the argument behind Theorem~\ref{thm:edge} no longer applies as it
stands. One can still follow the reduced state of the $d+1$ sites between the
edge and the readout, and this gives a correct law with $4^{d+1}$ hidden
states (Appendix~\ref{app:window}). That description, however, hides the
structure of the problem. The natural coordinates are the diagonals of the
circuit.

\begin{figure}[t]\centering
\resizebox{\columnwidth}{!}{% Light-cone coordinates of the brickwork.  x0 = 5, S = 4, sites 0..7.
% Drawn at its natural size (about 8 cm wide), so \resizebox{\columnwidth}
% scales it by a factor close to one and the text stays at \small.
\begin{tikzpicture}[x=0.92cm,y=1.2cm,>=Stealth,font=\small,
    ro0/.style={circle,fill=black,inner sep=2.3pt},
    ro1/.style={rectangle,draw=black,line width=0.9pt,fill=white,inner sep=2.6pt},
    ro2/.style={regular polygon,regular polygon sides=3,draw=black,line width=0.8pt,fill=black!45,inner sep=1.5pt}]
  \definecolor{lzero}{HTML}{e42536}
  \definecolor{lone}{HTML}{5790fc}
  \definecolor{conn}{HTML}{f89c20}
  % site wires
  \foreach \x in {0,...,7} {\draw[gray!45] (\x,0.35) -- (\x,4.75);}
  % background gates
  \foreach \h/\pty in {1/0,2/1,3/0,4/1} {
    \foreach \q in {0,...,6} {
      \pgfmathtruncatemacro{\ok}{mod(\q+\pty,2)==0 ? 1 : 0}
      \ifnum\ok=1
        \draw[gray!60,fill=gray!12,rounded corners=1pt] (\q-0.12,\h-0.17) rectangle (\q+1.12,\h+0.17);
      \fi
    }
  }
  % gates of L_0 (red) and L_1 (blue)
  \foreach \h in {1,2,3,4} {
    \pgfmathsetmacro{\qa}{5-\h}
    \pgfmathsetmacro{\qb}{7-\h}
    \draw[lzero,fill=lzero!18,rounded corners=1pt,thick] (\qa-0.12,\h-0.17) rectangle (\qa+1.12,\h+0.17);
    \draw[lone,fill=lone!18,rounded corners=1pt,thick] (\qb-0.12,\h-0.17) rectangle (\qb+1.12,\h+0.17);
  }
  % the two left-moving lines
  \draw[lzero,very thick] (5,0.35) -- (5,0.83)
    \foreach \h in {1,2,3,4} { -- ({5-\h+1},{\h-0.17}) -- ({5-\h},{\h+0.17}) } -- (1,4.75);
  \draw[lone,very thick] (7,0.35) -- (7,0.83)
    \foreach \h in {1,2,3,4} { -- ({7-\h+1},{\h-0.17}) -- ({7-\h},{\h+0.17}) } -- (3,4.75);
  % connectors c_0: right output of an L_0 gate, left input of the next L_1 gate
  \foreach \h in {1,2,3} {
    \draw[conn,very thick] ({6-\h},{\h+0.17}) -- ({6-\h},{\h+0.83});
  }
  % readouts
  \foreach \h in {1,2,3,4} {\node[ro0] at ({5-\h},{\h+0.5}) {};}
  \foreach \h in {1,2,3,4} {\node[ro1] at ({6-\h},{\h+0.5}) {};}
  \foreach \h in {1,2,3,4} {\node[ro2] at ({7-\h},{\h+0.5}) {};}
  % the diagonal cut through the d = 2 readout after layer 3
  \draw[dashed,thick] (2.25,2.0) -- (4.85,4.6);
  \node[anchor=west] at (4.85,4.6) {cut};
  % line labels at the top ends of the lines
  \node[lzero,anchor=south] at (1,4.75) {$L_0$};
  \node[lone,anchor=south] at (3,4.75) {$L_1$};
  % initial state along the bottom
  \node[anchor=north] at (5,0.30) {germ};
  \draw[decorate,decoration={brace,mirror,amplitude=4pt},thick] (-0.15,0.28) -- (4.15,0.28)
    node[midway,below=5pt] {$\mathds{1}/2$};
  \draw[decorate,decoration={brace,mirror,amplitude=4pt},thick] (5.85,0.28) -- (7.15,0.28)
    node[midway,below=5pt] {$\mathds{1}/2$};
  % half-step labels
  \foreach \h in {1,2,3,4} {\node[anchor=east] at (-0.3,\h) {$h=\h$};}
  % key, one row under the picture
  \node[ro0] (k0) at (0.6,-0.55) {};
  \node[anchor=west] at (k0.east) {\,readout $d=0$};
  \node[ro1] (k1) at (3.9,-0.55) {};
  \node[anchor=west] at (k1.east) {\,$d=1$};
  \node[ro2] (k2) at (5.7,-0.55) {};
  \node[anchor=west] at (k2.east) {\,$d=2$};
\end{tikzpicture}}
\caption{The brickwork in light-cone coordinates. The left-moving lines $L_0$
(red), the edge of the germ, and $L_1$ (blue) are joined by right-moving
connectors (orange). Information passes from $L_0$ to $L_1$ and never back. A
readout on the edge ($d=0$, circles) projects the co-moving qubit of $L_0$
itself. A readout one site inside ($d=1$, squares) projects the connector that
has just collided with that qubit, and a readout two sites inside ($d=2$,
triangles) projects the qubit of $L_1$. A right-moving cut (dashed) separates
the past of a record from its future and crosses one wire per line, which
bounds the memory of the record (Corollary~\ref{cor:ladder}).}
\label{fig:comoving}
\end{figure}
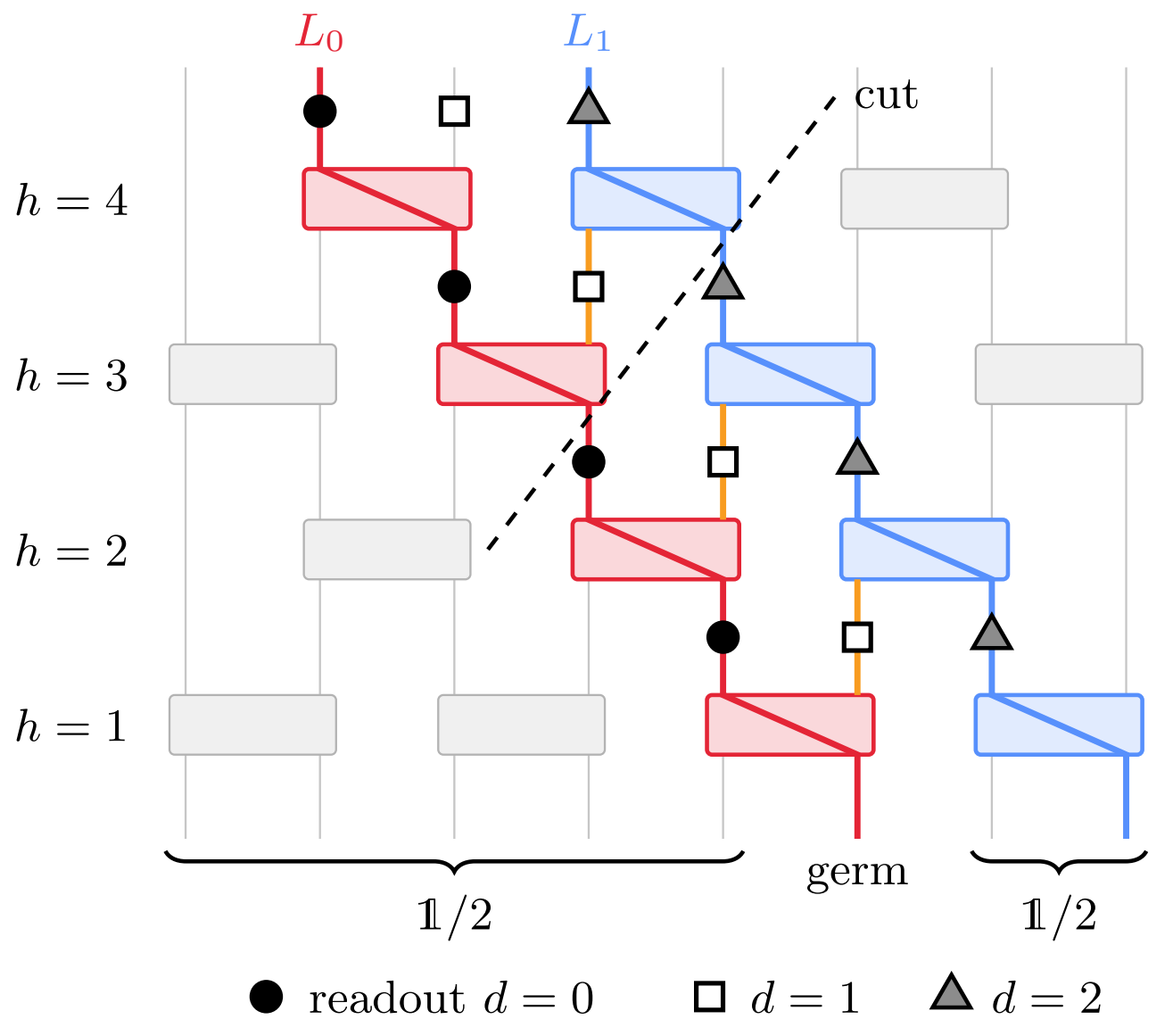

Every wire segment of a brickwork circuit runs either from the right input of
a gate to its left output or from its left input to its right output, and we
call the first kind left moving. The left-moving segments join into diagonal
lines $L_0,L_1,\dots$, where $L_0$ is the edge that carries the germ
(Fig.~\ref{fig:comoving}). In layer $h$ the gate of $L_i$ acts on the bond
$(x_0-h+2i,\,x_0-h+2i+1)$. Its left output $a_i$ continues along $L_i$, $2i$
sites inside the edge. Its right output $c_i$, $2i+1$ sites inside, enters
the gate of $L_{i+1}$ in the next layer as the left input, and we call it a
connector. The left input of $L_0$ comes from outside the cone. Information
thus flows along each line and from each line to the one behind it, but
never forward. Seen along its diagonals, the brickwork is a cascade.

\begin{theorem}[Co-moving frame]
\label{thm:comoving}
Let the readout sit $d$ sites inside the edge, at $x_0-h+d$, and let
$J=\lfloor d/2\rfloor$. Under the hypotheses of Theorem~\ref{thm:edge}, the
record is generated by the $J+1$ co-moving qubits $a_0,\dots,a_J$ alone. In
every layer the gate of $L_i$ acts on $(c_{i-1},a_i)$, with
$c_{-1}=\tfrac{\id}{2}$ a fresh site. The readout $K_m$ of Eq.~\eqref{eq:K}
acts on $a_J$ if $d$ is even and on the outgoing connector $c_J$ if $d$ is
odd, after which $c_J$ is discarded. Writing $\mathcal I_m$ for the resulting
linear map on the strip of co-moving qubits,
\begin{equation}
p(\bm m)=\Tr\big[\mathcal I_{m_S}\cdots\mathcal I_{m_1}(\rho_{\rm strip})\big],
\label{eq:hier}
\end{equation}
where $\rho_{\rm strip}$ places the germ on $a_0$ and $\tfrac{\id}{2}$ on
every other qubit.
\end{theorem}

A likelihood costs $O(S\,16^{J+1})$ operations, and the length of the chain
appears nowhere. For $d=0$ the strip is the single qubit $a_0$, and
Eq.~\eqref{eq:hier} reduces to Eq.~\eqref{eq:law}.

The proof (Appendix~\ref{app:comoving}) has two halves. Lines ahead of $L_0$
lie outside the forward cone of the germ and of every readout, so they feed
$L_0$ with maximally mixed qubits, just as in Theorem~\ref{thm:edge}. At
the other end, the connector $c_J$ leaves at maximal speed, while every later
readout sits on $L_J$ or on a connector leaving it. The connector thus never
enters the backward cone of a later readout, and trace preservation removes
it together with everything it touches further on. In site
coordinates the strip is simply the window of $d+1$ sites between the edge
and the readout. For the propagation of the reduced state alone, the
evolution of such a window is the channel of Ref.~\cite{SinghDooley}, and the
strip is the thin limit of the light-cone contraction schemes of
Refs.~\cite{FriasBanuls,KosBertiniProsen}. What these constructions do not
contain is the record, with the back-action of every measurement, and what
the co-moving picture implies for its memory.

How much does the record remember? Split it at some time into a past and a
future, and arrange the joint probabilities into a matrix
$p(\mathrm{past},\mathrm{future})$ whose rows are labelled by past words and
whose columns are labelled by future words. Its rank is the dimension of the
smallest linear model that generates the record~\cite{Jaeger}. We define the
memory of the record as the base-two logarithm of the largest rank over all
cuts. A sequence of independent bits has no memory, and a two-state Markov
chain has one bit.

\begin{corollary}[Memory ladder]
\label{cor:ladder}
For a readout $d$ sites inside the edge,
$\mathrm{rank}\,p(\mathrm{past},\mathrm{future})\le2^{d+1}$ for every cut.
The memory of the record is therefore at most $d+1$ bits, whatever its
length.
\end{corollary}

To see this, cut the network along a right-moving diagonal that passes
between the last readout of the past and the first readout of the future
(Fig.~\ref{fig:comoving}). The cut crosses each of $L_0,\dots,L_J$ once.
Every other wire it crosses carries only the identity, ahead of $L_0$ because
those qubits are maximally mixed and behind $L_J$ because nothing there is
ever read. Each crossed line can carry an operator from a four-dimensional
space, which bounds the rank by $4^{J+1}=2^{d+1}$ for odd $d$. For even $d$
the crossed wire of $L_J$ has just been projected by $K_m$ onto the span of
$\id$ and $Z$, and the bound becomes $2\cdot4^{J}=2^{d+1}$. The window
description, with its $4^{d+1}$ states, is thus the square of what the record
needs.

The bound is not a loose estimate. Almost every gate attains it.

\begin{proposition}[Generic saturation]
\label{prop:generic}
Fix $d$. If one two-qubit gate attains
$\mathrm{rank}\,p(\mathrm{past},\mathrm{future})=2^{d+1}$ for some cut of
some record length, then so does every gate outside a set of Haar measure
zero. Such a gate exists for every $d\le6$.
\end{proposition}

Every entry of $p(\mathrm{past},\mathrm{future})$ is a polynomial in the
entries of $U$ and their complex conjugates, so a minor of order $2^{d+1}$ is
a real-analytic function on the connected group $U(4)$. Such a function
either vanishes identically or vanishes only on a set of measure zero, and a
single gate with a nonzero minor decides which. We find such gates for every
$d\le6$ by constructing the reachable states of the strip directly, without
enumerating records, and the smallest retained singular value stays at least
six orders of magnitude above rounding error (Appendix~\ref{app:comoving}).
For these displacements the memory of the record is exactly $d+1$ bits for
almost every gate, and the states the strip can reach span exactly $2^{d+1}$
dimensions (Table~\ref{tab:ver}). We expect the same for every $d$, but we
have no witness construction that is uniform in $d$.

The ladder has a simple physical reading. On the edge the measurement acts on
the co-moving qubit itself. The projection wipes out everything the qubit
knew, the hidden state is the last outcome, and the record is the Markov
chain of Corollary~\ref{cor:chain}. One site inside, the qubit of $L_0$ is
never measured. In every layer it collides with a fresh qubit arriving from
outside the cone, and that qubit is measured as it leaves. This is a
collision model with indirect measurement~\cite{Ciccarello}, realized here
exactly inside a scrambling chain of arbitrary length. The unmeasured qubit
keeps a coherent memory of its entire history, and the record is not a Markov
chain of any finite order. For the Haar-random gate of our demonstrations,
outcomes older than the last four still change the probability of the next
outcome by up to $0.054$, although two bits of memory generate the record
exactly. We stress that $d$ is a property of the readout, for a fixed and
arbitrary gate. It has nothing to do with the hierarchies of gate conditions
that extend solvability into the cone for special
gates~\cite{YuWangKos,Rampp2026}. At every displacement we tested, the law
agrees with full density-matrix simulation to machine precision
(Table~\ref{tab:ver} and Appendix~\ref{app:window}).

\begin{figure}[t]
\includegraphics[width=\columnwidth]{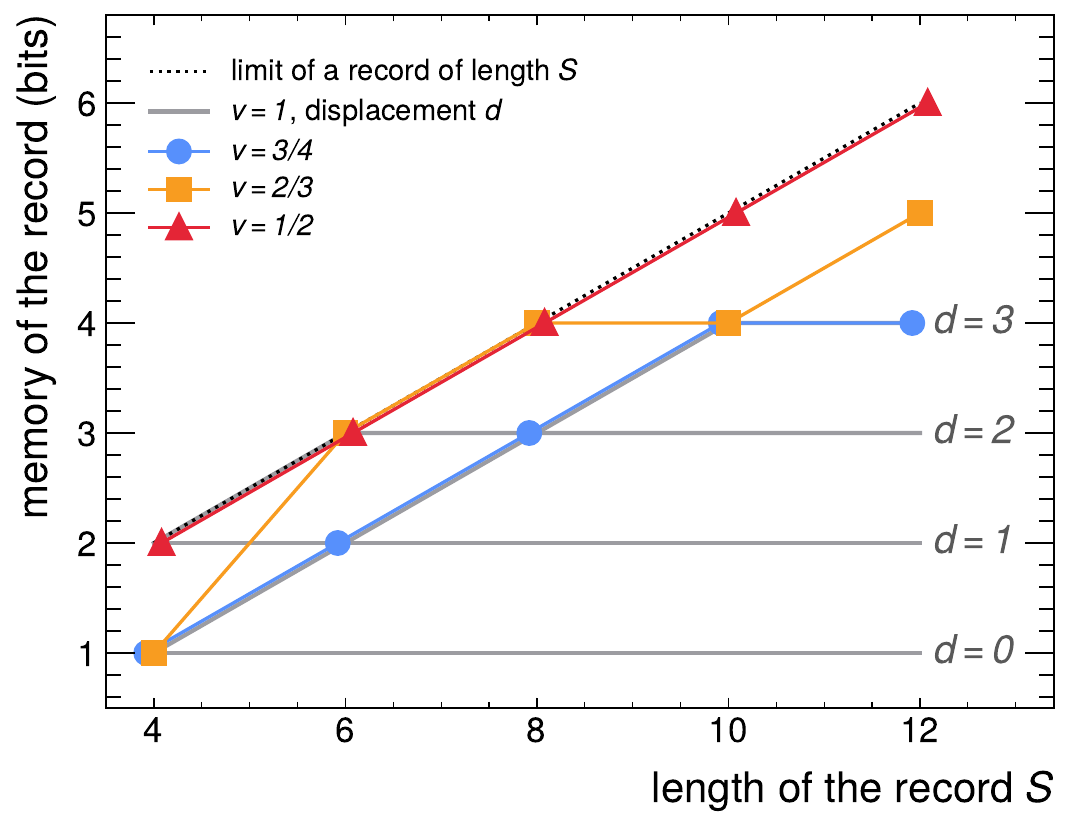}
\caption{Memory of the record, the base-two logarithm of the largest rank of
$p(\mathrm{past},\mathrm{future})$ over all cuts, as a function of the
length $S$ of the record, for a Haar-random gate. On rays of maximal speed the
memory saturates at $d+1$ bits and then stays there as the record grows (grey
lines). Below the maximal speed it grows with $S$. The dotted
line marks the largest memory a record of length $S$ can have. All values are
exact and follow from Theorem~\ref{thm:comoving}.}
\label{fig:ladder}
\end{figure}

\emph{Slower readouts.} A readout that moves at velocity $v<1$, at site
$x_0-\lceil vh\rceil$, has a displacement $d(h)=h-\lceil vh\rceil$ that grows
with time. The same strip with $J=\lfloor d(S)/2\rfloor$ gives its exact law,
and the cut argument shows that its memory can grow by at most $1-v$ bits per
half-step. For Haar-random gates it does grow, at every velocity we computed,
whereas on every ray of maximal speed it saturates at $d+1$ bits
(Fig.~\ref{fig:ladder}). The edge, in other words, is not the boundary of
solvability but the first rung of a ladder, and a record can become hard to
compute only if its readout falls behind the edge without bound.

\subsection{A witness of dual-unitarity}
\label{sec:witness}

At $d=1$ the germ and every readout lie strictly inside the light cone.
Dual-unitary circuits carry correlations only along the light-cone
edges~\cite{BKP2019}, so for a dual-unitary gate the $d=1$ record is exactly
uniform. In the co-moving picture, the qubit that leaves a dual-unitary
collision carries no information about the qubit it met, and the record has
no memory at all. On a device this null is spoiled by effects that have
nothing to do with the gate, such as relaxation toward $|0\rangle$ and
asymmetric readout. We therefore work with the germ-antisymmetrized record
\begin{equation}
W_h=\tfrac12\big(\langle m_h\rangle_{+\bm r}-\langle m_h\rangle_{-\bm r}\big).
\label{eq:witness}
\end{equation}
Record probabilities are linear in $\rho_0$, so $W_h$ is the response of the
record to the germ alone, and every contribution that does not depend on the
germ cancels exactly. Local unitaries preserve dual-unitarity, so coherent
single-qubit errors cannot produce a signal either. We have checked that
$W_h$ vanishes to machine precision for a dual-unitary gate under
depolarizing noise and amplitude damping on the rail, relaxation of every
qubit in every time slice, and arbitrary readout assignment errors
(Table~\ref{tab:ver}). A nonzero $W_h$ at $d=1$ therefore certifies that the
executed two-qubit dynamics has left the dual-unitary manifold, and
Theorem~\ref{thm:comoving} predicts its value for any given gate.

\section{Checks against exact simulation}
\label{sec:verify}

Every statement above was checked against full density-matrix simulation of
the system together with the readout, and each check was paired with a
control designed to fail if the claim were wrong (Table~\ref{tab:ver}). The
controls fail at the percent level, by amounts that depend on the gate, so we
quote the values for the ensembles we used and not as universal constants.
Two entries deserve a comment. The late-twirl checks simulate the executed
schedule, with damping in every time slice including the readout windows, so
the object that is verified is the object that runs. And one control fails
silently. The untransposed law passes a test with a $z$-polarized germ to
machine precision and is exposed only by transverse germs.

\begin{table}[t]\footnotesize
\caption{Verification against full density-matrix simulation of system plus
readout, maxima over the stated ensembles. Executed-schedule checks include
damping in every time slice. The controls in the lower block are designed to
falsify the claims and must fail, one of them silently. DU denotes the
dual-unitary gate.}
\label{tab:ver}
\begin{ruledtabular}
\begin{tabular}{lc}
Test & error \\ \hline
Eq.~\eqref{eq:law}, 5 Haar gates $\times$ 4 germs & $1.3\times10^{-15}$\\
Eq.~\eqref{eq:law}, inhomogeneous in time & $1.4\times10^{-16}$\\
Eq.~\eqref{eq:chain} against Eq.~\eqref{eq:law} & $2.8\times10^{-16}$\\
$n$-independence, $n=8\ldots12$ & $6.1\times10^{-16}$\\
Thm.~\ref{thm:noise}(i,ii), unital depolarizing $p=0.5$ off ray & $3.1\times10^{-16}$\\
Thm.~\ref{thm:noise}(i), non-unital $T_1$, $\gamma=0.3$, behind ray & $1.4\times10^{-17}$\\
Thm.~\ref{thm:comoving}, window form, $d=1,2,3$ ($4^{d+1}$ states) & $5.1\times10^{-16}$\\
Thm.~\ref{thm:comoving}, co-moving against window form, $d\le3$ & $2.4\times10^{-17}$\\
Thm.~\ref{thm:comoving}, $d=4,5$, executed schedule & $6.2\times10^{-17}$\\
Thm.~\ref{thm:comoving}, $d=1,2,3$, $T_1$ $\gamma\le0.15$ everywhere & $3.6\times10^{-16}$\\
Thm.~\ref{thm:comoving}, every gate of the chain different, $d=2$ & $6.9\times10^{-17}$\\
Thm.~\ref{thm:comoving}, readout at $v=\tfrac12$ and $v=0$ & $8.3\times10^{-17}$\\
Prop.~\ref{prop:generic}, rank at $d=0\ldots6$, Haar gates & $2^{d+1}$ exactly\\
Thm.~\ref{thm:noise}(iii), executed schedule, $T_1$, $\gamma=0.15$ & $8.3\times10^{-17}$\\
Thm.~\ref{thm:noise}(iii), Eq.~\eqref{eq:noisylaw}, $T_1$ $\gamma=0.2$ everywhere & $5.6\times10^{-17}$\\
Thm.~\ref{thm:noise}(iii), kernel under injected idle, $\gamma=0.3$ & $1.1\times10^{-16}$\\
$W_h$, $d=1$, DU gate, noise and readout error & $8.3\times10^{-17}$\\ \hline
Wrong sublattice & $2.5\times10^{-2}$\\
Eq.~\eqref{eq:law} at readout $+1$ inside cone & $2.4\times10^{-2}$\\
Thm.~\ref{thm:comoving} with $d=1$ applied to $d=0$ data & $2.4\times10^{-2}$\\
Untransposed law, transverse germ & $3.1\times10^{-2}$\\
Untransposed law, $z$ germ (fails silently) & $6.9\times10^{-17}$\\
$t=0$ twirl schedule, $T_1$ $\gamma=0.05$ all slices & $1.6\times10^{-2}$\\
\end{tabular}
\end{ruledtabular}
\end{table}

\section{Demonstrations on a cloud quantum processor}
\label{sec:hardware}

Unless stated otherwise, the demonstrations ran on \textit{ibm\_fez}, a
156-qubit superconducting processor of IBM Quantum accessed through the
cloud, on linear chains of physical qubits with the germ on the sublattice
required by Theorem~\ref{thm:edge}. We call these runs demonstrations,
since the device was not under our control. No error
mitigation was used in any run, and the interleaved jobs used no
dynamical decoupling and no twirling by the runtime. The maximally mixed
background was
realized by randomizing the background sites from shot to shot. In the
recorded rails the dual-unitary gate was compiled from an exact three-CNOT
template~\cite{VatanWilliams}, and in the interleaved demonstrations every
gate was compiled from its matrix by the transpiler. The processors, the
dates, the job identifiers, the calibration data at the time of each run and
the software settings are documented in Appendix~\ref{app:methods}.

\subsection{Recorded rails obey the Markov structure}
\label{sec:rails}

Our first test uses twelve rails recorded before the present analysis, at
chain lengths $n=12$ to $54$ and depths $S=4$, $6$ and $8$, with
$2\times10^4$ shots each. Eleven are driven by the dual-unitary gate and one
by a Haar-random gate. For each rail we fit Eq.~\eqref{eq:chain}, extended by
asymmetric readout assignment errors, to the full distribution of the $2^S$
outcomes, once with a kernel common to all steps and once with a separate
kernel for each step (Table~\ref{tab:rails}). The common-kernel chain fails,
with a pooled $\chi^2/\mathrm{dof}=9.52$. The step-dependent chain fits, with
a pooled $\chi^2/\mathrm{dof}=1.37$. It is also the correct null hypothesis,
since each step of a rail runs on a different pair of physical qubits with
its own gate and readout errors. The deepest rail is the most demanding test,
and it is described at $\chi^2/\mathrm{dof}=1.44$ on $206$ degrees of
freedom. Two of the twelve rails stay above $3$.

The test also tells noise on the rail apart from defective hardware. A
single-qubit rotation planted deliberately on the rail changes the kernel but
not the Markov structure (Sec.~\ref{sec:law}), and indeed the planted records
fit as well as the controls measured on the same chains, with
$\chi^2/\mathrm{dof}$ between $0.6$ and $3.9$ against $1.6$ to $3.8$. Chains
that contain a qubit with a stuck readout, recorded on
\textit{ibm\_marrakesh}, fail even the step-dependent test, at
$\chi^2/\mathrm{dof}$ between $110$ and $201$.

\begin{table}[t]\footnotesize
\caption{Markov-structure test of Corollary~\ref{cor:chain} on twelve rails
recorded on \textit{ibm\_fez} with $2\times10^4$ shots each. Reduced $\chi^2$
of Eq.~\eqref{eq:chain}, extended by asymmetric readout assignment errors,
against the full distribution of $2^S$ outcomes, with one kernel common to
all steps and with one kernel per step. DU denotes the dual-unitary gate.
Degrees of freedom are given in brackets.}
\label{tab:rails}
\begin{ruledtabular}
\begin{tabular}{lcclcc}
Rail & $n$ & $S$ & gate & common & per step\\ \hline
R1 & 12 & 4 & DU & 23.44 (10) & 0.75 (6)\\
R2 & 12 & 4 & DU & 4.60 (10) & 1.65 (6)\\
R3 & 12 & 4 & DU & 3.09 (10) & 1.52 (6)\\
R4 & 20 & 4 & DU & 7.56 (10) & 1.27 (6)\\
R5 & 20 & 4 & DU & 3.41 (10) & 1.38 (6)\\
R6 & 20 & 4 & DU & 7.89 (10) & 3.08 (6)\\
R7 & 20 & 4 & Haar & 8.25 (10) & 1.16 (6)\\
R8 & 20 & 6 & DU & 1.52 (57) & 0.61 (49)\\
R9 & 50 & 4 & DU & 3.09 (10) & 4.89 (6)\\
R10 & 50 & 8 & DU & 12.49 (213) & 1.44 (206)\\
R11 & 54 & 4 & DU & 7.67 (10) & 1.75 (6)\\
R12 & 54 & 4 & DU & 8.46 (10) & 0.27 (6)\\
\hline
pooled & & & & 9.52 & 1.37\\
\end{tabular}
\end{ruledtabular}
\end{table}

The rail means themselves are the same function of $h$ at $n=12$, $20$, $50$
and $54$. Random Pauli errors injected on every off-ray site at rates up to
$p=0.5$ leave them unchanged within statistics, with weighted slopes of
$+0.012\pm0.014$, $+0.013\pm0.017$, $+0.001\pm0.018$ and $-0.007\pm0.018$
per unit of $p$ at the four steps, as Theorem~\ref{thm:noise} predicts for
unital noise. Pauli injection cannot probe the non-unital half of part (ii),
and that is the purpose of the demonstration described in Sec.~\ref{sec:idle}.

\subsection{An interleaved demonstration of noise protection and of the witness}
\label{sec:interleaved}

The remaining predictions were tested in a single job on one chain of twenty
qubits, with the germ at chain position $x_0=17$ and the rail region on
physical qubits $147$ to $142$, whose reported $T_1$ times lay between $102$
and $155\,\mu$s. Background sites were prepared by a Hadamard gate followed
by a mid-circuit measurement, which produces $\tfrac{\id}{2}$ exactly in each
shot. We ran eight configurations, each with the germ at $+\bm r$ and at
$-\bm r$ and $16\,384$ shots per germ sign, and interleaved the $128$
resulting circuit blocks in time so that drift cannot distinguish one
configuration from another. The late twirl was realized exactly, by
enumerating all $4^S$ Pauli assignments of the incoming sites with equal
numbers of shots. Two-qubit gates were compiled from their matrices by the
transpiler, and we checked that no compiled circuit contained a SWAP and that
every record bit was measured on the intended physical qubit.

The acceptance criteria were set before submission. With the late twirl,
every component of the idle-induced kernel shift had to satisfy $|z|<3$, and
with the twirl at $t=0$ at least one component had to exceed $|z|=5$. The
dual-unitary witness had to satisfy $|z|<3$ at every step, and the Haar
witness had to exceed $|z|=5$ at the first step. Every late-twirl
configuration had to pass the step-dependent Markov test at
$\chi^2/\mathrm{dof}<2$. All criteria were met.

All six configurations at $d=0$ pass the step-dependent Markov test
(Table~\ref{tab:campaign}), and the agreement with the noiseless law is
close. The first-step germ-odd mean of the dual-unitary gate is $0.6784$,
against a predicted $0.6803$, and the measured kernel contractions
$s_h=0.569$, $0.570$ and $0.547$ lie within ten percent of the noiseless
value $0.607$. A fit of Eq.~\eqref{eq:noisylaw} to the dual-unitary data
gives a rail depolarizing strength of $1.2\%$, a post-readout relaxation of
$3.9\%$ and readout assignment errors below $0.5\%$.

\begin{table}[t]\footnotesize
\caption{Markov-structure test of the interleaved demonstration at $d=0$, with
$3.3\times10^4$ shots per configuration and both germ signs fitted jointly
with a shared kernel. The common kernel has $24$ degrees of freedom and the
per-step kernel $20$.}
\label{tab:campaign}
\begin{ruledtabular}
\begin{tabular}{lcc}
Configuration & common & per step\\ \hline
dual-unitary, twirl at $t=0$ & 1.89 & 1.29\\
dual-unitary, late twirl & 1.63 & 1.05\\
Haar, twirl at $t=0$ & 1.98 & 0.69\\
Haar, late twirl & 2.55 & 1.28\\
Haar, twirl at $t=0$, idle injected & 22.76 & 0.73\\
Haar, late twirl, idle injected & 1.60 & 1.21\\
\end{tabular}
\end{ruledtabular}
\end{table}

\subsection{Injected idle relaxation}
\label{sec:idle}

\begin{figure}[t]
\includegraphics[width=\columnwidth]{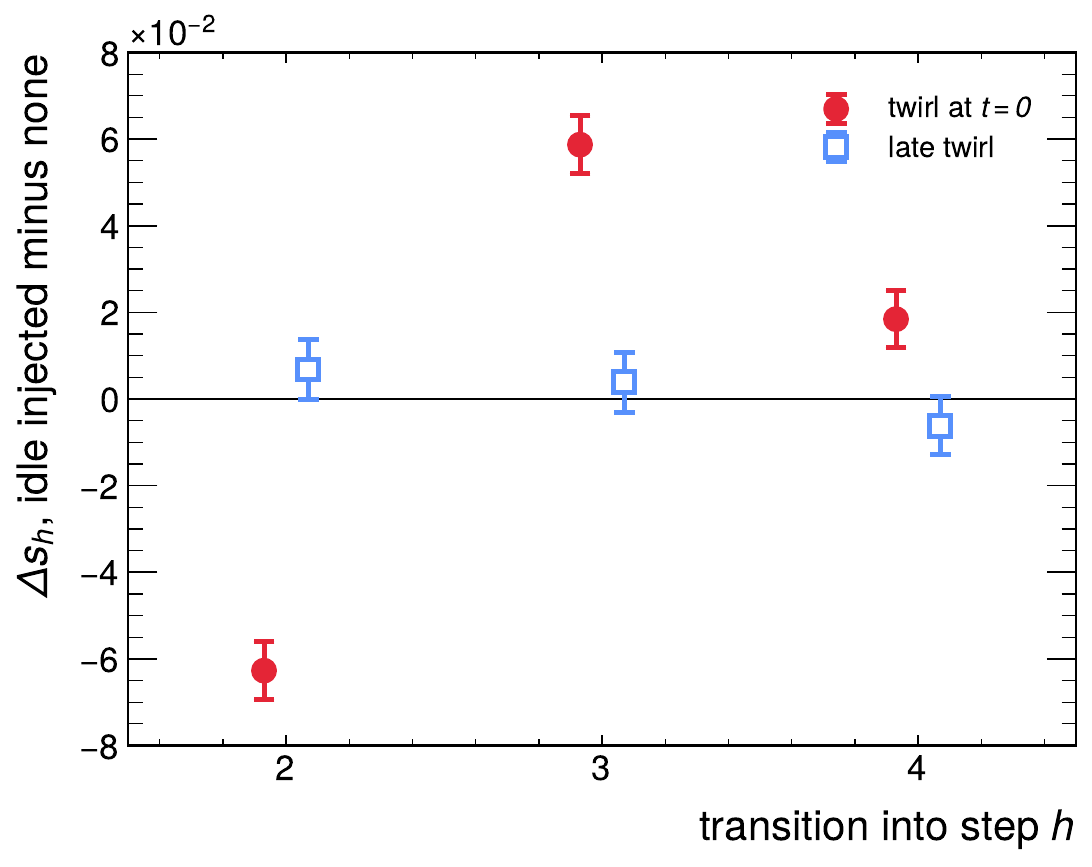}
\caption{Part (iii) of Theorem~\ref{thm:noise} on hardware. Shift of the
kernel contraction $s_h$ of the monitored chain, Eq.~\eqref{eq:chain},
caused by an idle period of $\tau=50\,\mu$s injected on every qubit, for the
Haar-random gate. Filled circles show the twirl at $t=0$ and open squares the
late twirl. Both configurations ran on the same chain in one interleaved job,
with $3.3\times10^4$ shots each. Error bars are bootstrap standard errors.}
\label{fig:idle}
\end{figure}

To test part (ii) of Theorem~\ref{thm:noise} and its remedy, we inserted an
idle period of $\tau=50\,\mu$s on every qubit of the chain after the
background preparation. This relaxes each incoming site by
$1-e^{-\tau/T_1}\approx0.3$ before it reaches the ray. With the twirl at
$t=0$, this non-unital noise ahead of the ray shifts the kernel contractions
by $\Delta s_2=-0.063$ and $\Delta s_3=+0.059$, at $9.5\sigma$ and
$8.4\sigma$ (Fig.~\ref{fig:idle}). With the late twirl, on the same chain and
in the same job, no component of the kernel moves by more than $1.5\sigma$,
as part (iii) predicts. Without injection the two protocols differ little, by
at most $2.2\sigma$ for the dual-unitary gate and by $3.6\sigma$ in one
kernel component for the Haar gate, consistent with the near-immunity of
dual-unitary gates discussed in Sec.~\ref{sec:noise}. The injected relaxation
changes the kernel but not the Markov structure. The $t=0$ record with
injected idle still fits a step-dependent chain at $\chi^2/\mathrm{dof}=0.73$
and fails a common-kernel chain at $22.8$.

\subsection{Measuring the executed gate}
\label{sec:witnesshw}

\begin{figure}[t]
\includegraphics[width=\columnwidth]{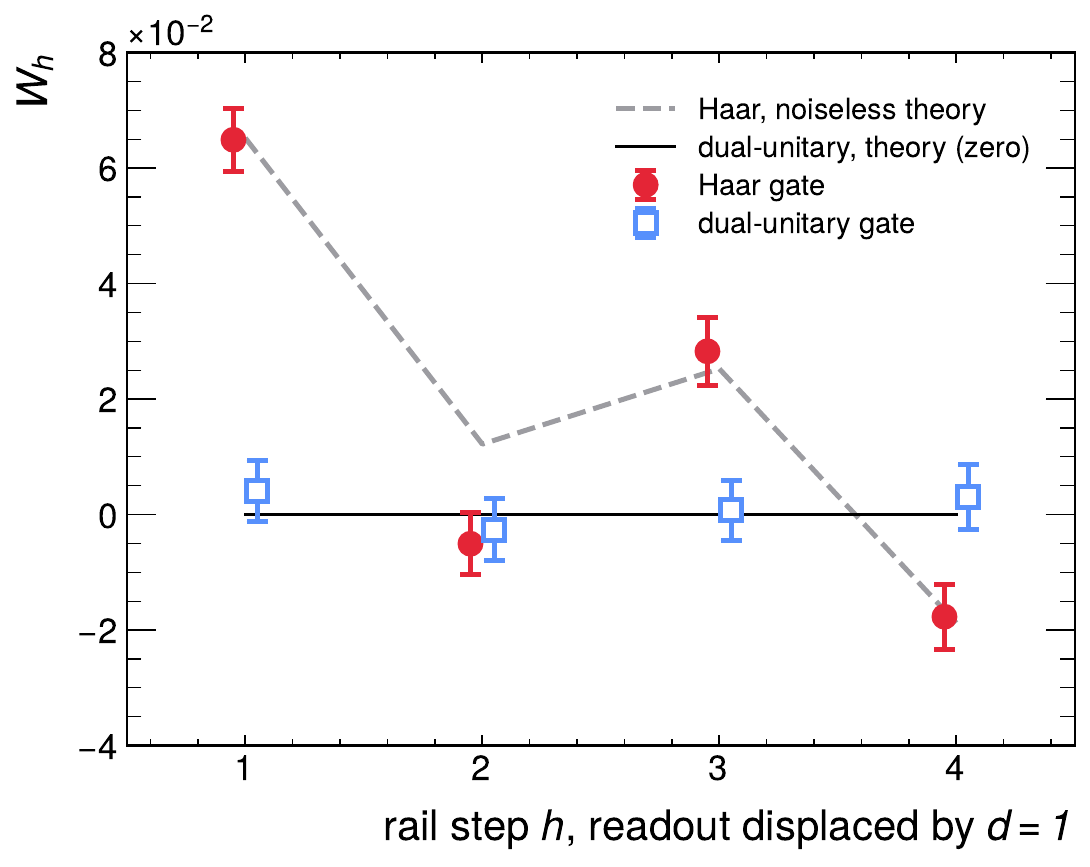}
\caption{The germ-antisymmetrized record $W_h$ of Eq.~\eqref{eq:witness},
measured one site inside the edge ($d=1$) on \textit{ibm\_fez}. Open squares
show the dual-unitary gate, for which the theory predicts exactly zero.
Filled circles show a Haar-random gate on the same qubits, with the noiseless
prediction of Theorem~\ref{thm:comoving} as a dashed line. We used
$3.3\times10^4$ shots per gate. Error bars are bootstrap standard errors.}
\label{fig:witness}
\end{figure}

At $d=1$ the dual-unitary gate gives $W_h=(0.004,-0.003,0.001,0.003)$ with a
standard error of $0.0055$, null at every step to within $0.7\sigma$
(Fig.~\ref{fig:witness}). The Haar-random gate on the same qubits gives
$W_1=0.065$, at $11.8\sigma$, against a noiseless prediction of $0.065$. At
steps $1$, $3$ and $4$ the Haar signal agrees with the noiseless law of
Theorem~\ref{thm:comoving} to within $0.6\sigma$, and at step $2$ it differs
by $3.1\sigma$. The dual-unitary witness at the same step is null, so the
deviation is not a readout effect.

The null is in fact a measurement of the executed gate. We displace the
nonlocal coefficients of the gate away from the dual-unitary point,
$e^{-i[(\pi/4-\delta_a)XX+(\pi/4-\delta_b)YY+JZZ]}$ with the local dressing
held fixed, propagate each candidate through Eq.~\eqref{eq:hier} with the
noise calibrated on the $d=0$ data of the same chain, and compare with the
measured $W_h$ (Appendix~\ref{app:witness}). At $95\%$ confidence,
$|\delta|\le0.011$\,rad for a common displacement $\delta_a=\delta_b$ and
$|\delta|\le0.010$\,rad for a differential one, $\delta_a=-\delta_b$. The
corresponding dual-unitarity defect is at most $0.042$, compared with $0.797$
for the Haar gate. The bound needs neither tomography nor a fitted baseline,
and single-qubit noise cannot fake a violation.

\subsection{Memory of the displaced record}
\label{sec:memory}

Theorem~\ref{thm:comoving} predicts more than the means of the $d=1$ record.
It predicts that the record of a Haar-random gate remembers, and that the
edge record and the record of the dual-unitary gate do not. The simplest
observable that sees the difference is the order-two memory
\begin{align}
\delta(h,a)={}&P(m_h{=}{+}\,|\,m_{h-1}{=}a,\,m_{h-2}{=}{+})\nonumber\\
&-P(m_h{=}{+}\,|\,m_{h-1}{=}a,\,m_{h-2}{=}{-}),
\label{eq:memory}
\end{align}
which vanishes identically for a Markov chain and for a sequence of
independent bits. As for the witness, we use its germ-antisymmetrized part
$\Delta=\tfrac12[\delta_{+\bm r}-\delta_{-\bm r}]$. The reason is that a
readout assignment error turns the chain of Corollary~\ref{cor:chain} into a
hidden chain, which has a memory of its own, but that memory does not depend
on the germ. For the assignment errors of this chain, up to $1.4\%$, exact
simulation gives a germ-even memory of $0.019$ and a germ-odd part below
$5\times10^{-4}$.

\emph{Discovery.} We first looked for memory in the records of the
interleaved job after it had run, outside its acceptance criteria. For the
Haar gate at $d=1$ these records give $\Delta(3,+)=0.051\pm0.008$, a
$6.5\sigma$ effect, against $0.052$ from Theorem~\ref{thm:comoving} with no
free parameter, and $\chi^2=50.6$ against zero on four components. The
dual-unitary gate at $d=1$ and both gates at $d=0$ give $\chi^2$ between
$2.0$ and $7.7$ (Appendix~\ref{app:memfirst}).

\emph{Registered test.} An effect found after the fact needs independent
confirmation. Before any further data existed, we specified a second
demonstration, its observable, its error estimate, its noise model and five
acceptance criteria, and recorded them together with a checksum of the
analysis code (Appendix~\ref{app:memreg}). Nine configurations were interleaved in one job
(\texttt{daop43opqrnc739aancg}), namely the dual-unitary gate, the Haar gate
of the first job (gate A) and a second Haar-random gate (gate B), each at
$d=0$, $1$ and $2$, all with the late twirl and both germ signs, with
$3.3\times10^4$ shots per configuration and germ sign, $589\,824$ in all.
Gate B was chosen in advance for its large predicted memory. The chain
consists of the twenty qubits of the first job traversed in the opposite
direction, which puts the germ on qubit $110$ and the edge rail on qubits
$109$, $118$, $129$ and $130$, none of which played these roles before. The
prediction is the law of Theorem~\ref{thm:comoving} with a single noise model
for the chain, fitted jointly to the three $d=0$ records of the same job. No
parameter is fitted to a displaced record.

With $Q=18.47$, the $99.9\%$ point of the $\chi^2$ distribution with four
degrees of freedom, the criteria read as follows. (K1) The dual-unitary
records at $d=1$ and $2$ give $\chi^2<Q$ against zero, and the three edge
records give $\chi^2<Q$ against the law. (K2) Gate A at $d=1$ reproduces
$\Delta(3,+)$ above $5\sigma$, with $\chi^2<Q$ against the law. (K3, K4)
Gate B differs from zero at $\chi^2>Q$ at both $d=1$ and $d=2$, exceeds
$5\sigma$ in the component that the noiseless law predicts to be largest, and
agrees with the law at $\chi^2<Q$, and gate A at $d=2$ agrees with the law as
well. (K5) A step-dependent two-state Markov chain is rejected at
$p<10^{-6}$ for both Haar gates at $d=1$, each with a $\chi^2/\mathrm{dof}$
more than $2.5$ times the largest value among the edge records and the
dual-unitary record at $d=1$.

\begin{figure}[t]
\includegraphics[width=\columnwidth]{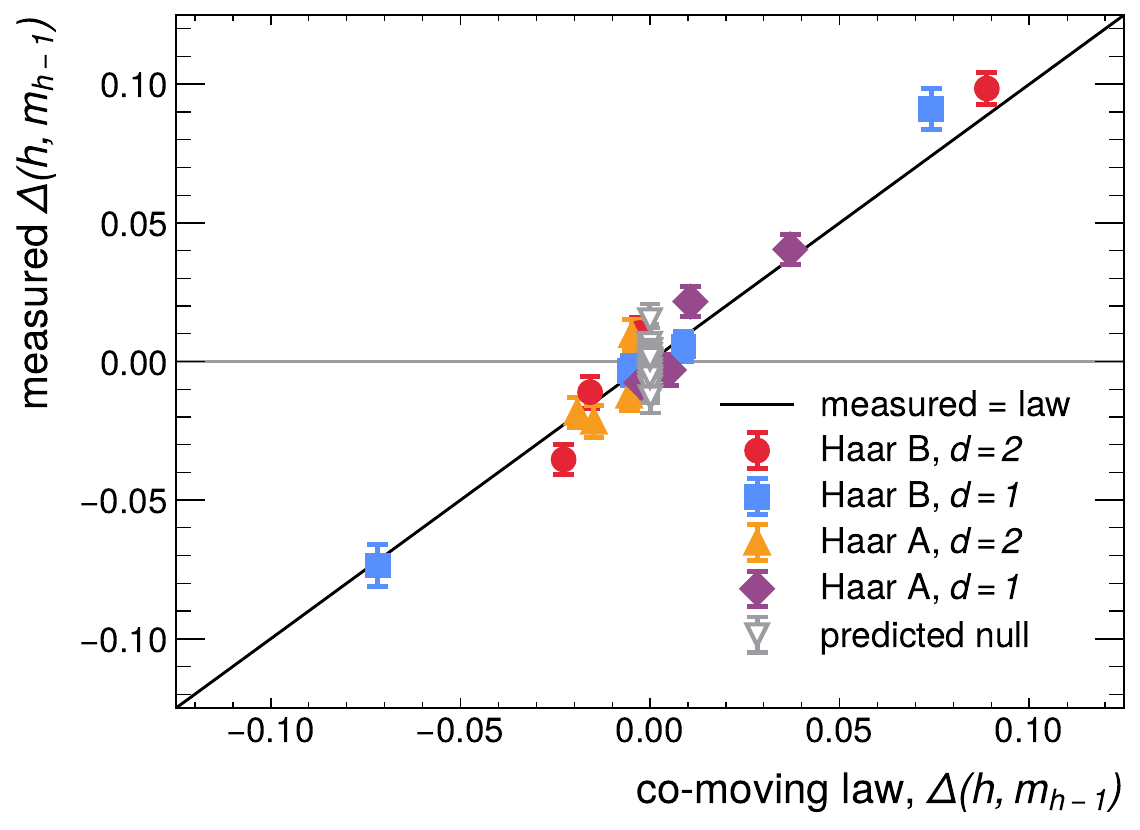}
\caption{Memory of the record in the registered demonstration on
\textit{ibm\_fez}. The measured germ-antisymmetrized order-two memory
$\Delta(h,m_{h-1})$ of Eq.~\eqref{eq:memory} is plotted against the law of
Theorem~\ref{thm:comoving}, with the chain noise fitted to the $d=0$ records
of the same job, for all $36$ components of the nine configurations. Open
triangles mark the five records predicted to carry no germ-odd memory, the
dual-unitary gate at $d=0$, $1$ and $2$ and both Haar gates at $d=0$. We used
$3.3\times10^4$ shots per configuration and germ sign. Error bars are
binomial standard errors.}
\label{fig:memory}
\end{figure}

All five criteria are met (Fig.~\ref{fig:memory} and
Table~\ref{tab:memory}). Gate A at $d=1$ gives $\Delta(3,+)=0.040\pm0.006$,
at $7.3\sigma$, against a predicted $0.037$. Gate B gives $-0.074\pm0.007$
and $+0.091\pm0.007$ at $d=1$, against $-0.072$ and $+0.074$, and
$0.098\pm0.006$ at $d=2$, at $17.6\sigma$, against $0.089$. The five records
predicted to carry no germ-odd memory give $\chi^2$ between $1.0$ and $12.4$.
The Markov chain describes the edge records and the dual-unitary record at
$d=1$ at $\chi^2/\mathrm{dof}$ between $0.93$ and $1.39$, and it is rejected
for the two Haar records at $d=1$ at $11.6$ and $30.3$ on $12$ degrees of
freedom, which corresponds to $p=7\times10^{-24}$ and $2\times10^{-70}$.

The chain was noisier than in the first job, with a fitted depolarizing
strength of $5.4\%$ per layer, and the noise matters. Against the noiseless
law, gate B gives $\chi^2=54$ at $d=1$ and $154$ at $d=2$, against $5.9$ and
$14.3$ for the law with the chain noise. Over all $36$ components the law
with the chain noise gives $\chi^2=60.5$, somewhat more than the number of
components, which reflects the limits of a five-parameter noise model
(Appendix~\ref{app:memreg}). One and two sites inside the edge, then, the
record of a scrambling gate is no longer a Markov chain, and the co-moving
qubits account for the size of the departure.

\emph{Repetitions.} The registration provided for repetitions on other
processors, each analysed on its own against the same criteria. The same
nine configurations were submitted to \textit{ibm\_marrakesh} and
\textit{ibm\_kingston} within fifteen minutes of the job on
\textit{ibm\_fez} (Appendix~\ref{app:memreg} and Table~\ref{tab:repeat}).
Every prediction that does not involve the noise model holds on all three
processors. The fifteen records predicted to carry no germ-odd memory give
$\chi^2<Q$, the twelve displaced Haar records differ from zero at $\chi^2$
between $30$ and $623$, gate A at $d=1$ exceeds $5\sigma$ in $\Delta(3,+)$
each time, and the step-dependent Markov chain is rejected for both Haar
gates at $d=1$ at $p<10^{-19}$ on every processor. What does not carry over
unchanged is the agreement in size under one five-parameter noise model per
chain. On \textit{ibm\_kingston} gate B at $d=1$ gives
$\chi^2_{\rm law}=21.1$, just above $Q$, so that K3 fails while the other
four criteria are met. On \textit{ibm\_marrakesh} the noise model fails
already on the edge records to which it is fitted, at
$\chi^2/\mathrm{dof}=10.5$, although the step-dependent Markov chain
describes the same records at $0.76$ to $0.84$. There K3 and K4 fail, and
gate B at $d=1$ agrees instead with the noiseless law, at $\chi^2=12.9$.
Against the law with the chain noise, eight of the twelve displaced Haar
records lie below $Q$. The memory is therefore reproduced on three
processors, while a quantitative prediction of its size is only as good as
the model of the noise of the chain.

\begin{table}[t]\footnotesize
\caption{The registered memory demonstration on \textit{ibm\_fez}. For each
record, $\chi^2$ of the germ-antisymmetrized order-two memory on its four
components against zero ($\chi^2_0$) and against the law of
Theorem~\ref{thm:comoving} with the chain noise ($\chi^2_{\rm law}$), the
significance of the component that the noiseless law predicts to be largest,
and $\chi^2/\mathrm{dof}$ of a step-dependent two-state Markov chain fitted
to each germ sign ($12$ degrees of freedom). The registered threshold is
$Q=18.47$. DU denotes the dual-unitary gate.}
\label{tab:memory}
\begin{ruledtabular}
\begin{tabular}{lcccc}
record & $\chi^2_0$ & $\chi^2_{\rm law}$ & largest & Markov\\ \hline
DU, $d=0$ & 0.9 & 1.0 & & 0.97\\
DU, $d=1$ & 3.5 & 3.5 & & 0.93\\
DU, $d=2$ & 5.4 & 5.4 & & \\
Haar A, $d=0$ & 12.3 & 12.4 & & 1.39\\
Haar A, $d=1$ & 71.6 & 7.3 & $7.3\sigma$ & 11.63\\
Haar A, $d=2$ & 34.0 & 9.5 & $3.4\sigma$ & \\
Haar B, $d=0$ & 1.2 & 1.2 & & 0.99\\
Haar B, $d=1$ & 251.8 & 5.9 & $9.8\sigma$ & 30.30\\
Haar B, $d=2$ & 356.5 & 14.3 & $17.6\sigma$ & \\
\end{tabular}
\end{ruledtabular}
\end{table}

\subsection{Locating a faulty gate}
\label{sec:diagnosis}

Causality also localizes faults. The correlator $\langle m_1m_{1+k}\rangle$
involves only the gates of steps $2,\dots,k+1$, so the first lag at which it
deviates points to a single gate. On \textit{ibm\_marrakesh} one correlator
exceeded its noiseless value, the opposite of what decoherence normally does,
and the analysis pointed to the gate of step~2. The corresponding qubit
returned outcome $1$ in $6.5\%$ of shots, and
$\langle m_1m_2\rangle\simeq\langle m_1\rangle$, measured as $0.4505$
against $0.4332$, confirmed a nearly constant bit. The calibration record of
the provider at the time listed this qubit with a readout assignment error
of $2.6\%$ and $T_1=188\,\mu$s, inside the cuts used to select the chain,
so the calibration did not reveal the fault. The two other runs on the same
chain, submitted eight minutes earlier and 73 minutes later, single out the
same qubit, and it later exceeded the device's own readout-error threshold. Faults comparable to the baseline error of a chain
require a paired reference measured in the same job. With planted rotations
of $0.8$\,rad, the paired comparison identifies the planted step on both
processors. The plant at step~2 moves $\langle m_2\rangle$ by $-0.209$ and
$-0.200$, at about $30\sigma$, while $\langle m_1\rangle$, which causality
protects, stays within $1.6\sigma$, and a control compared with itself flags
nothing.

\section{Discussion}
\label{sec:discussion}

The record law is a likelihood, and therefore also a sampler, so records on
the edge carry no sampling advantage. Exactness and classicality have the
same origin. Moving the readout into the cone does not destroy exactness. It
raises the memory by one bit per site and the cost by a factor of sixteen for
every two sites, in the displacement and not in $n$. For generic gates a
monitored record can be hard to compute only if the readout moves strictly
slower than the edge, so that it falls behind through a number of lines that
grows with time, adding at most $1-v$ bits of memory per half-step
(Fig.~\ref{fig:ladder}). The co-moving frame thus turns the question of where
hardness lives into a count of lines, and it provides exact references along
the way. A record at $v=\tfrac12$ and depth $S=12$ involves four co-moving
lines, and its law is computed exactly on eight qubits, however long the
chain.

For special gates, solvability reaches further into the cone, through
hierarchies of gate conditions~\cite{YuWangKos,Rampp2026} and through
constructed circuits whose solvable influence matrices are
non-Markovian~\cite{Pickering}. Space-time duality maps sub-luminal monitored
records onto unitary correlation functions of rotated
circuits~\cite{IppolitiKhemani,LuGrover}, which is the natural setting in
which to ask when a record becomes hard. As an instrument, the record law
complements the benchmarking of mid-circuit measurements~\cite{Govia,Hothem}.
Those protocols characterize the measurement operation itself, whereas the
record law provides an exact reference for a scrambling circuit as a whole.

The scope of the results should be stated plainly. All exact statements
require the sites that join the edge to arrive maximally mixed and
uncorrelated. The infinite-temperature background provides this, and the
late twirl enforces it whatever the background, but the twirl is an average
over shots. It does not describe a chain quenched from a pure state, in which
the bath ahead of the edge has a memory of its own. After such quenches the
temporal entanglement of influence matrices in chaotic circuits follows a
volume law in time~\cite{FZB}, and we see no reason to expect a ladder of
fixed height there. On the hardware side, the Markov tests reach depth
$S=8$, while the memory demonstrations use $S=4$, which is too short for a
direct test of the rank and leaves the order-two memory as the natural probe.
The noise model behind the displaced predictions has five parameters and
treats the chain as homogeneous. Fitted to the edge records it gives
$\chi^2/\mathrm{dof}=2.72$ on \textit{ibm\_fez}, $1.94$ on
\textit{ibm\_kingston} and $10.5$ on \textit{ibm\_marrakesh}, and the
predicted size of the memory meets the registered threshold in all four
displaced Haar records on the first processor, in three on the second and
in one on the third. The memory itself appears on all three, but a
prediction of its size requires a model of the noise of the chain, which
the present one only approximates. Finally, Proposition~\ref{prop:generic}
is established for $d\le6$ only.

Two extensions follow directly from the construction. The cut argument does
not use the identity of the gates, and with the late twirl in place local
noise on the strip can at most double the bound at even $d$. The rank of
$p(\mathrm{past},\mathrm{future})$, or the decay of its singular values,
would therefore test the memory ladder on hardware without any noise model,
given records long enough to estimate it. The ladder also indicates where to
look for records that are hard to predict, namely behind the edge at
sub-maximal speed, where the number of co-moving lines grows with time.

\section{Conclusion}

The output of a quantum processor that measures qubits mid-circuit is a
classical record, and for a scrambling circuit that record is in general as
hard to predict as the circuit itself. We have shown that near the light cone
this is not so. Causality alone makes the record on the edge exactly
Markovian, for arbitrary gates and at any system size, and seen from a frame
that moves with the edge, a scrambling chain is a memoryless bath for the few
qubits the observer carries. The memory of the record is quantized, one bit
on the edge and at most one more for every site of displacement into the
cone, and generic gates saturate the bound.

For the validation of quantum hardware this provides what was missing in
the regime of dynamic circuits, an exact reference for a scrambling circuit
with mid-circuit measurements whose cost does not grow with the device,
together with a precise account of the noise it is blind to, a remedy at no
gate cost for the one channel it is not blind to, and a witness of the
dual-unitarity of the executed gate that single-qubit noise cannot fake. For
the theory of monitored many-body dynamics it locates where the output can
become hard to predict, namely behind the edge at sub-maximal speed, and it
shows that the collision models of open-system theory arise exactly inside
generic scrambling circuits.

On a 156-qubit superconducting processor accessed through the cloud, twelve
independently recorded rails obey the predicted Markov structure, and
non-unital noise injected ahead of the ray corrupts the record only when the
late twirl is absent. The witness bounds the nonlocal coefficients of the
compiled dual-unitary gate to $0.011$\,rad, and a registered demonstration
finds the memory of the co-moving qubits one and two sites inside the edge,
for two scrambling gates, at the predicted values, and none where none is
predicted. Registered repetitions on two further processors find the memory
and its absence in the same places, and they show that its predicted size
is only as reliable as the model of the noise of the chain.

\begin{acknowledgments}
The author acknowledges support from the Helmholtz
Association HGF (Germany), the Hamburgische Investitions- und F\"orderbank
(IFB) (Germany), and the Ministry of Science, Research and Culture of the
State of Brandenburg within the Center for Quantum Technologies and
Applications (CQTA) (Germany). The hardware runs used the IBM Quantum
processors, accessed through DESY.
\end{acknowledgments}

\emph{Data and code availability.} The hardware records of every run in
Table~\ref{tab:platform}, the calibration records and job identifiers of
Appendix~\ref{app:methods}, the circuits and results of every identified
job, the protocol of the registered memory demonstration, the
exact-simulation code and the analysis code that regenerate every number and
figure in this work are available from the author and will be deposited in a
public repository upon publication.

\appendix

\section{The two channels and why the transpose matters}
\label{app:transpose}

Write the single-gate Pauli channel of $U$ as $\mathcal M(\sigma_a)=
\tfrac12\Tr_1[U^\dagger(\sigma_a\otimes\id)U]$, a real $4\times4$ matrix in
the basis $\{\id,X,Y,Z\}$. For any single-qubit operator $X_1$ and any state
$\rho_2$,
\begin{equation}
\Tr\!\big[X_1\,\Tr_2[U(\tfrac{\id}{2}\otimes\rho_2)U^\dagger]\big]
=\Tr\!\big[\tfrac12\Tr_1[U^\dagger(X_1\otimes\id)U]\,\rho_2\big].
\end{equation}
Hence $\mathcal M$ is the Heisenberg adjoint of the Schr\"odinger map
$\Phi(\rho)=\Tr_2[U(\tfrac{\id}{2}\otimes\rho)U^\dagger]$, which carries the
state of the right site to the left site with a maximally mixed partner, and
in an orthonormal Pauli basis $A=\mathcal M^{\mathsf T}$. The record
propagates states, not observables, so $A$ and not $\mathcal M$ is the
transfer matrix. The identity is elementary, but confusing the two matrices
is an easy mistake to make and a hard one to detect. The block probed by a
$z$-polarized germ is symmetric, so a test with such a germ cannot tell the
two apart. Any reimplementation should therefore be validated with
transverse germs and over an ensemble of gates. Over $12$ Haar-random gates
the untransposed law fails by a median of $8.1\times10^{-3}$, but by as
little as $2.0\times10^{-4}$ for individual gates, so a single test with a
transverse germ is not a reliable discriminator either. The same caution
applies to every control in Table~\ref{tab:ver2}. Their failure magnitudes
span more than two decades across Haar gates and should not be quoted as
constants.

\section{Proof of Theorem~\ref{thm:edge}}
\label{app:recordlaw}

The theorem follows from a statement about the state just before each edge
gate. The edge gate of half-step $h+1$ acts on the pair
$(x_0-h-1,\,x_0-h)$, where $x_0-h$ is the rail site read at half-step $h$
and $x_0-h-1$ is the incoming site. We claim that the reduced state of this
pair, conditioned on the outcomes $m_1,\dots,m_h$, is exactly
$\tfrac{\id}{2}\otimes\rho_{x_0-h}$, with the incoming site as the first
factor and $\rho_{x_0-h}$ the conditional state of the rail site.

\emph{Proof.} Write the global state after the readout of half-step $h$,
conditioned on the outcomes so far and not normalized, as
$2^{-n}(c\,\id+O)$, with $c$ the probability of these outcomes and $O$ a
traceless operator. Initially $c=1$ and $O=\bm r\cdot\bm\sigma^{(x_0)}$ is
supported on the germ. A brickwork layer conjugates $O$ by a unitary and
enlarges its support by at most one site in each direction. A readout of site
$j$ with outcome $m$ replaces $c\,\id+O$ by $P_m(c\,\id+O)P_m$, with
$P_m=\tfrac12(\id+mZ_j)$, and adds support only on $j$. Every readout so far
took place on a site $x_0-j$ with $j\le h$, inside the forward cone of the
germ, so after half-step $h$ the support of $O$ is contained in
$[x_0-h,\,x_0+h]$. Now take the partial trace onto
$T=\{x_0-h-1,\,x_0-h\}$. Every Pauli string in $O$ with a nontrivial factor
outside $T$ traces to zero, and no string has a factor on $x_0-h-1$, which
lies outside the cone. What survives is $c\,\id/4$ plus strings supported on
$x_0-h$ alone, which is $\tfrac{\id}{2}\otimes\rho_{x_0-h}$. $\square$

The edge gate is the only gate of layer $h+1$ that touches $x_0-h-1$, so the
conditional state of the next rail site is
$\Tr_2[U(\tfrac{\id}{2}\otimes\rho_{x_0-h})U^\dagger]$, whose Bloch vector is
obtained by applying $A_{h+1}$ to that of $\rho_{x_0-h}$
(Appendix~\ref{app:transpose}). The right output of the gate falls behind the
ray and never influences a later readout. The readout then applies
Eq.~\eqref{eq:K}, and composing the $S$ steps gives Eq.~\eqref{eq:law}. The
parity hypothesis is what allows a single gate per half-step to carry the
ray. If $x_0$ and $x_0-1$ do not share a gate at $h=1$, the measured sites no
longer trace the edge and the law fails (Table~\ref{tab:parity}).

\section{Proof of Corollary~\ref{cor:chain}}
\label{app:cor1}

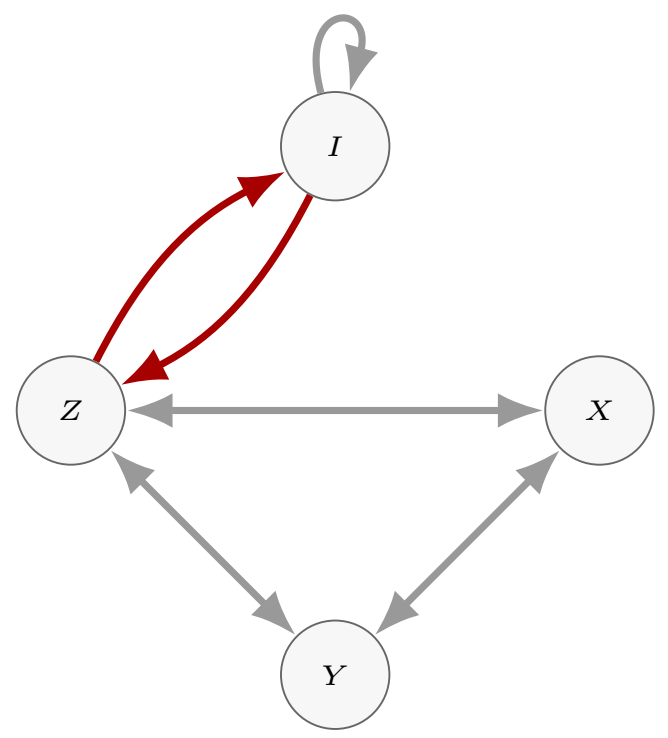
\begin{figure}[h]\centering
\resizebox{0.5\columnwidth}{!}{% The four Pauli components of the ray.  A_h is trace preserving, so no weight
% flows from X, Y, Z into I, and for a unitary gate with a maximally mixed
% incoming leg it is unital, so I maps to I.  K_m exchanges weight between I
% and Z only and removes X and Y.
\begin{tikzpicture}[>=Latex, line join=round,
  live/.style={red!65!black, line width=1.5pt}, tiny/.style={font=\small}]
\foreach \i/\ang/\nm in {0/90/{I},1/0/X,2/270/Y,3/180/Z}{
  \node[circle,draw=black!60,fill=black!3,minimum size=8mm,inner sep=0pt]
    (s\i) at (\ang:1.95) {$\scriptstyle \nm$};
}
\foreach \i/\j in {1/2,2/3,3/1}{
  \draw[black!40,<->,line width=1.5pt] (s\i) -- (s\j);
}
\path (s0) edge[loop above, black!40, ->, line width=1.5pt, min distance=7mm] (s0);
\draw[live,->] (s3) to[bend left=18] (s0);
\draw[live,->] (s0) to[bend left=18] (s3);
\end{tikzpicture}}
\caption{The path of Fig.~\ref{fig:dag} as a walk on the four Pauli
components of the ray. Each edge gate applies $A_h$ (grey), which fixes $I$
and mixes $X$, $Y$ and $Z$, and each readout applies $K_{m_h}$ (red), which
keeps only $I$ and $Z$. After the first readout every walk stays on
$\{I,Z\}$, which is the content of Corollary~\ref{cor:chain}. Non-unital
noise on the rail adds an edge from $I$ to $Z$, the bias $t_h$. The
probability of a record is the total weight of the walk, a product of
$4\times4$ matrices however many qubits are present.}
\label{fig:digraph}
\end{figure}

Write the vector of the ray after the gate of half-step $h$ as $u=A_hv$. The
readout applies $K_m$, and for $m=\pm1$
\begin{equation*}
K_mu=\tfrac12\,(u_{\id}+m\,u_Z)\,(1,0,0,m)^{\mathsf T}.
\end{equation*}
The prefactor is the probability of the outcome and $(1,0,0,m)$ is the Bloch
vector of $|m\rangle$, so after the first readout the hidden state is the
last outcome itself. The next gate maps $(1,0,0,m_{h-1})$ to a vector whose
$Z$ component is $t_h+s_hm_{h-1}$, with $t_h=[A_h]_{Z\id}$ and
$s_h=[A_h]_{ZZ}$, so that $m_h$ occurs with probability
$\tfrac12[1+m_h(t_h+s_hm_{h-1})]$. Only the first readout sees the full germ
vector, through $\mu_1=[A_1v_0]_Z$, and multiplying the factors gives
Eq.~\eqref{eq:chain} (Fig.~\ref{fig:digraph}). For a unitary gate with a
maximally mixed incoming site the edge propagator is unital and $t_h=0$. A
single-qubit channel acting on the rail between readouts replaces $A_h$ by
its composition with that channel, which changes $(t_h,s_h)$ but not the form
of the law. A readout assignment error acts on the recorded bit after the
state update and turns the chain into a hidden two-state chain. The identity
holds numerically to $2.8\times10^{-16}$ (Table~\ref{tab:ver}).

\section{Proof of Theorem~\ref{thm:noise}}
\label{app:dichotomy}

Let a channel $\mathcal E$ act on site $q$ at half-step $h$.

(i) Suppose $q>x_0-h$. Information propagates by at most one site per
half-step, while the readouts move left by exactly one site per half-step. A
perturbation at $(q,h)$ can reach site $q-(h'-h)$ at half-step $h'$, and
$q-(h'-h)>x_0-h'$ for all $h'>h$. It never meets the ray, and the record is
invariant for arbitrary $\mathcal E$, unital or not.

(ii) Suppose $q<x_0-h$. This site joins the rail at half-step $h'=x_0-q>h$ as
the incoming site of an edge gate, and the proof of
Appendix~\ref{app:recordlaw} requires it to arrive as $\tfrac{\id}{2}$. The
region ahead of the ray is in the maximally mixed state, which unitaries and
unital channels leave unchanged, so unital channels are invisible. For
necessity, note that a non-unital $\mathcal E$ produces an incoming Bloch
vector $\bm z\neq0$, on which the edge propagator depends linearly,
$A(\bm z)=A(0)+\sum_az_aA'_a$, with $A'_a\neq0$ for generic $U$.

(iii) For any state $\rho$ of the chain,
$\tfrac14\sum_{P\in\{\id,X,Y,Z\}}P_q\rho P_q=
\tfrac{\id_q}{2}\otimes\Tr_q\rho$. A uniformly random Pauli operator applied
to site $q$ therefore erases its polarization and all of its correlations,
exactly, on average over shots. Applied to each incoming site just before its
edge gate, the twirl restores the hypothesis of Appendix~\ref{app:recordlaw}
regardless of what happened to the site before. Single-qubit channels on the
rail compose with $A_h$ as in Appendix~\ref{app:cor1}, and an assignment
error replaces $K_m$ by $K^{(q)}_m$, which gives Eq.~\eqref{eq:noisylaw}. A
further channel between a readout and the next edge gate is absorbed into the
next propagator, so the law also covers relaxation during the mid-circuit
measurement itself, which lasts about $1\,\mu$s. $\square$

\emph{Consequences for hardware.} Amplitude damping is non-unital, so part
(ii) applies precisely to the $S$ qubits of the rail that the ray has not yet
reached. These qubits idle from $t=0$ through several mid-circuit
measurements and arrive polarized. Full density-matrix simulation gives a
record deviation of $6.0\times10^{-3}$ at $\gamma=0.05$ and
$1.2\times10^{-2}$ at $\gamma=0.10$, first order in $\gamma$
(Table~\ref{tab:ver2}). The naive repair, feeding a damped state
$\tfrac12(\id+z_hZ)$ into $A$, is itself wrong at order $\gamma$, because the
incoming sites are also correlated with one another through the gates they
passed through while idle. Its residual is $3.6\times10^{-3}$ at
$\gamma=0.02$. The late twirl removes the problem at the cost of moving $S$
single-qubit gates later in the schedule. Simulating the executed schedule
with $T_1$ damping on every qubit in every time slice, layers and readout
windows alike, gives $8.1\times10^{-16}$ at $\gamma=0.05$ and
$8.3\times10^{-17}$ at $\gamma=0.15$, whereas the same schedule with the
twirl left at $t=0$ deviates from any such law by $1.6\times10^{-2}$ and
$4.3\times10^{-2}$. The verification runs on an intermediate representation
of the schedule that is also what is emitted to Qiskit, so the object that is
checked is the object that runs (\texttt{runbook\_patch.py}).

\section{Proofs of Theorem~\ref{thm:comoving}, Corollary~\ref{cor:ladder}
and Proposition~\ref{prop:generic}}
\label{app:comoving}

\emph{Light-cone coordinates.} Label the gate on the bond
$(x_0-h+2i,\,x_0-h+2i+1)$ in layer $h$ as $G_i^h$. Under the parity
hypothesis these are all the gates of layer $h$, with $i$ running over the
integers allowed by the chain. The left output of $G_i^h$ is site
$x_0-h+2i$, which in layer $h+1$ is the right member of the bond
$(x_0-h-1+2i,\,x_0-h+2i)$ and hence the right input of $G_i^{h+1}$. The right
output of $G_i^h$ is site $x_0-h+2i+1$, the left member of the bond of
$G_{i+1}^{h+1}$. The gates $G_i^1,G_i^2,\dots$ thus form a left-moving
line $L_i$ threaded by one qubit $a_i$, and the only coupling between lines
is the connector $c_i$ from $G_i^h$ to $G_{i+1}^{h+1}$. An operator inserted
on $L_i$ can spread only to lines of index $i$ and higher, and one inserted on
$c_i$ only to lines of index $i+1$ and higher. The germ is the right input of
$G_0^1$ and lies on $L_0$.

\emph{Theorem~\ref{thm:comoving}.} Let the readouts sit at displacement $d$,
with $J=\lfloor d/2\rfloor$. For even $d$ they act on $a_J$ after each gate
of $L_J$, and for odd $d$ on $c_J$. (a) Lines of negative index receive only
maximally mixed inputs, because neither the germ nor any readout can spread
to them, and a unitary maps $\tfrac{\id}{4}$ to itself. The left input
$c_{-1}$ of every gate of $L_0$ is therefore $\tfrac{\id}{2}$ and
uncorrelated with the strip. (b) Lines of index larger than $J$ are never
read and never couple back, so by trace preservation the trace over them can
be taken as soon as $c_J$ leaves $L_J$. What remains is the evolution of
$a_0,\dots,a_J$ and of the connectors in flight between them, which is
Eq.~\eqref{eq:hier}. The argument holds gate by gate for inhomogeneous
circuits and for arbitrary single-qubit channels on the strip. With the late
twirl it holds for arbitrary noise ahead of $L_0$, and noise behind $L_J$ is
invisible by (b). For a readout trajectory whose displacement $d(h)$ never
exceeds $D$, the argument is unchanged with $J=\lfloor D/2\rfloor$, the
readout acting on whichever wire of the strip sits at displacement $d(h)$.
$\square$

\emph{Corollary~\ref{cor:ladder}.} The right outputs of the gates
$G_0^{h-J},G_1^{h-J+1},\dots,G_J^{h}$ lie on one right-moving diagonal
$R_h$. Consider the cut that runs between $R_k$ and $R_{k+1}$. It crosses
each line $L_i$ once, on the segment between two consecutive gates, and it
crosses no connector. For even $d$ the readout $m_k$ sits on the crossed
segment of $L_J$, and the cut is placed after it. The readouts
$m_1,\dots,m_k$ then lie on one side and $m_{k+1},\dots,m_S$ on the other.
By (a) the wires of lines of negative index carry $\tfrac{\id}{2}$ from the
past side, and by (b) the wires of lines of index above $J$ are contracted
with the identity on the future side, so both contribute rank one. The record
distribution is a contraction of a past tensor and a future tensor over the
operator spaces of the $J+1$ crossed wires, which bounds the rank of
$p(m_1\ldots m_k,\,m_{k+1}\ldots m_S)$ by $4^{J+1}$. For even $d$ the crossed
wire of $L_J$ has just been acted on by $K_m$, whose range is spanned by
$\id$ and $Z$, and the bound improves to $2\cdot4^{J}$. In both cases it
equals $2^{d+1}$. For a trajectory with $d(h)\le D$ the same cut gives
$4^{\lfloor D/2\rfloor+1}$. Table~\ref{tab:ladder} lists the ranks found
numerically. $\square$

\begin{table}[t]
\caption{Memory of the record in bits, the base-two logarithm of the largest
rank of $p(\mathrm{past},\mathrm{future})$ over all cuts of a record of
length $S=12$, which can show at most $6$ bits. Two Haar-random gates and a
dual-unitary gate, computed from Theorem~\ref{thm:comoving} by
\texttt{lightcone\_comoving.py}.}
\label{tab:ladder}
\begin{ruledtabular}
\begin{tabular}{lcccc}
ray & bound & Haar & Haar & dual-unitary\\ \hline
$v=1$, $d=0$ & 1 & 1 & 1 & 1\\
$v=1$, $d=1$ & 2 & 2 & 2 & 0\\
$v=1$, $d=2$ & 3 & 3 & 3 & 1\\
$v=1$, $d=3$ & 4 & 4 & 4 & 0\\
$v=1$, $d=4$ & 5 & 5 & 5 & 1\\
$v=3/4$ & 4 & 4 & 4 & 1\\
$v=2/3$ & 6 & 5 & 5 & 1\\
$v=1/2$ & 8 & 6 & 6 & 1\\
\end{tabular}
\end{ruledtabular}
\end{table}

\emph{Proposition~\ref{prop:generic}.} Each record probability is a
contraction of copies of $U$ and of $U^{*}$ with fixed tensors, and hence a
polynomial in the real and imaginary parts of the entries of $U$. A minor of
order $2^{d+1}$ of $p(\mathrm{past},\mathrm{future})$ is a
real-analytic function $f$ on $U(4)$, which is a connected real-analytic
manifold. If $f$ does not vanish identically, its zero set has Haar measure
zero. The hypothesis provides a gate with $f\neq0$ for a suitable minor, and
Corollary~\ref{cor:ladder} bounds the rank from above. $\square$

The witnesses are computed by \texttt{ladder\_saturation.py}, which needs no
enumeration of records. It builds the states that the strip can reach after a
past of length $d+4$, discarding at each step every state that is a linear
combination of states already kept. Nothing is lost by this pruning, because
the maps are linear. For the Haar gates tested the reachable states span
exactly $2^{d+1}$ dimensions. The script then evaluates all future words of
length $d+3$ on each kept state. Every kept state is the state after one
definite past word, so the resulting matrix is, up to a rescaling of its
rows, a submatrix of $p(\mathrm{past},\mathrm{future})$ for records of length
$2d+7$, and its rank bounds that of the full matrix from below.
Table~\ref{tab:witness} gives the smallest of its first $2^{d+1}$ singular
values relative to the largest. It stays far above the rounding error of
$10^{-16}$, while a singular value number $2^{d+1}+1$ does not exist because
the matrix has only $2^{d+1}$ rows. For $d\le6$ the memory of the record is
therefore exactly $d+1$ bits for almost every gate.

\begin{table}[t]\footnotesize
\caption{Witnesses for Proposition~\ref{prop:generic}. Number of linearly
independent reachable states and rank of the matrix of future-word
probabilities evaluated on them, with its smallest singular value relative to
the largest. Haar-random gates with seeds $4$ and $11$ of
\texttt{lightcone.haar\_u}.}
\label{tab:witness}
\begin{ruledtabular}
\begin{tabular}{ccccc}
$d$ & $2^{d+1}$ & reachable & rank & $\sigma_{\min}/\sigma_{\max}$ (seed 4, 11)\\ \hline
0 & 2 & 2 & 2 & $1.7\times10^{-1}$, $2.0\times10^{-1}$\\
1 & 4 & 4 & 4 & $3.8\times10^{-4}$, $2.2\times10^{-3}$\\
2 & 8 & 8 & 8 & $9.2\times10^{-6}$, $5.0\times10^{-6}$\\
3 & 16 & 16 & 16 & $5.3\times10^{-6}$, $1.1\times10^{-6}$\\
4 & 32 & 32 & 32 & $6.1\times10^{-8}$, $8.2\times10^{-9}$\\
5 & 64 & 64 & 64 & $2.3\times10^{-7}$, $1.7\times10^{-7}$\\
6 & 128 & 128 & 128 & $1.2\times10^{-10}$ (seed 4 only)\\
\end{tabular}
\end{ruledtabular}
\end{table}

\emph{Markov order.} A finite rank does not make the record a Markov chain of
finite order. At $d=1$ the hidden state is the unmeasured qubit of $L_0$,
which no finite string of outcomes determines. For the Haar gate of the
interleaved demonstration, the largest change in the probability of the next
outcome caused by the history before the last $k$ outcomes is $0.081$,
$0.081$, $0.056$, $0.054$ and $0.039$ for $k=1,\dots,5$, at $d=1$ in records
of length $S=10$, and zero to machine precision at $d=0$.

\section{The window form of the displaced-readout law}
\label{app:window}

The same record can be described in site coordinates. This is how it was
first derived, and it provides an independent check of
Theorem~\ref{thm:comoving}. Let the readout sit at $p_h=x_0-h+d$. A site at
$(p,h)$ influences the readout at half-step $h'>h$ only if
$|p_{h'}-p|\le h'-h$. Since $p_{h'}+h'$ is constant along a ray of maximal
speed, the backward cone of the whole record is bounded on the right by
$p\le x_0-h+d$ at half-step $h$. Intersecting it with the forward cone
$[x_0-h,\,x_0+h]$ of the germ leaves the window
$W_h=[x_0-h,\,x_0-h+d]$ of $d+1$ sites. Everything to the left of the window
is maximally mixed and uncorrelated with it, and everything to the right is
causally disconnected from the record. For the propagation of the reduced
state alone, the evolution of such a window is the channel $\Phi_{d+1}$ of
Ref.~\cite{SinghDooley}, independent of $n$. The new ingredient is its
interleaving with the non-unital readout update.

\begin{figure}[t]\centering
\includegraphics[width=\columnwidth]{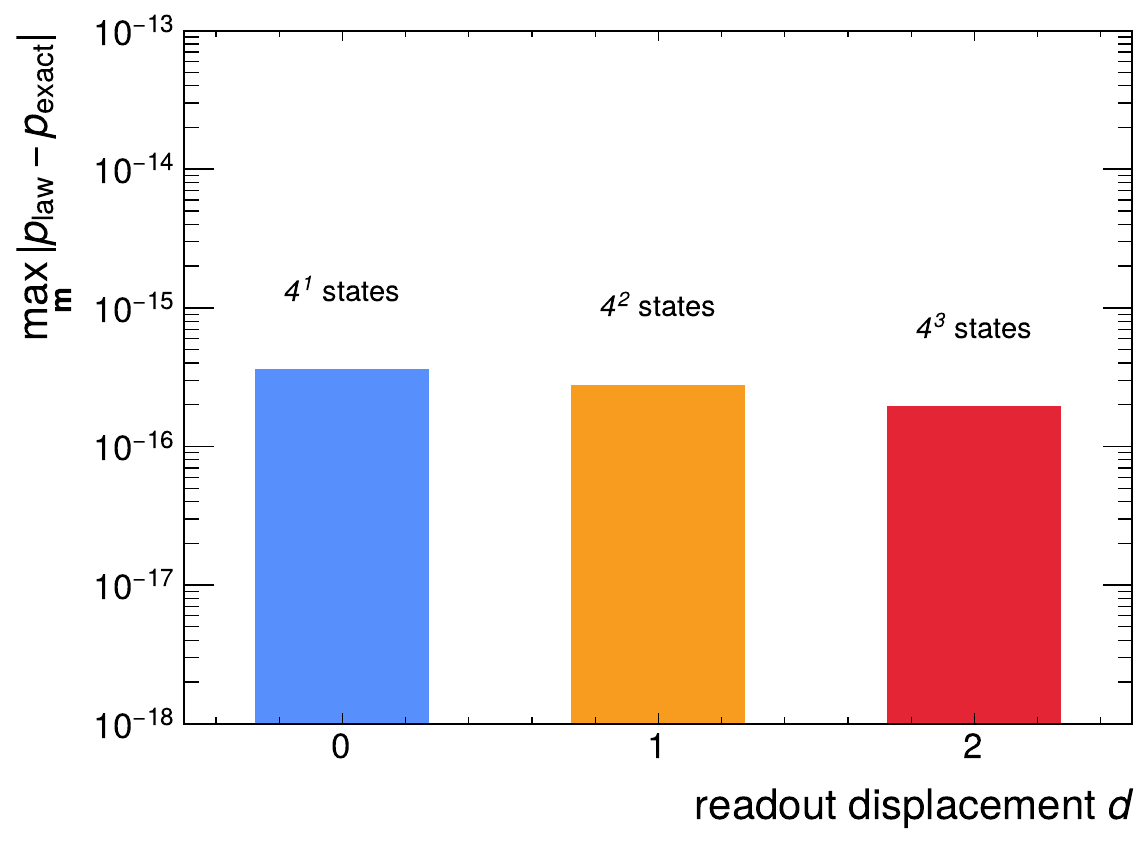}
\caption{Theorem~\ref{thm:comoving} against full density-matrix simulation.
Largest deviation between the displaced-readout law and exact simulation of
system plus readout, for a generic gate at $d=0$, $1$ and $2$, with
$4^{d+1}$ hidden states in the window form.}
\label{fig:hierprec}
\end{figure}

\begin{figure}[t]
\includegraphics[width=\columnwidth]{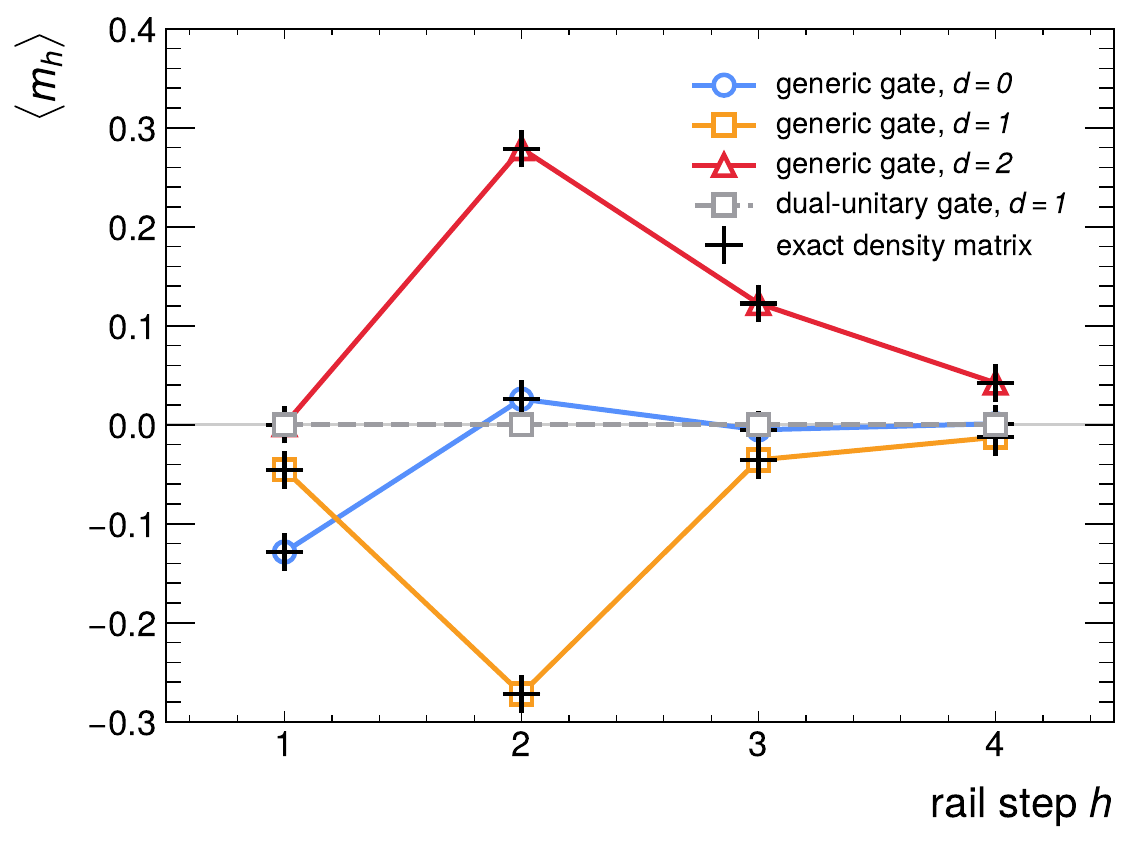}
\caption{Displaced rail means for a generic gate at $d=0$, $1$ and $2$.
Lines show the law of Theorem~\ref{thm:comoving} and crosses show full
density-matrix simulation. The dual-unitary null at $d=1$ is included for
comparison. The law agrees with the simulation to machine precision
(Table~\ref{tab:ver} and Fig.~\ref{fig:hierprec}).}
\label{fig:hier}
\end{figure}

The transfer from $W_h$ to $W_{h+1}$ takes three steps. Adjoin a maximally
mixed site on the left, apply the gates of layer $h+1$ whose support lies
inside $[x_0-h-1,\,x_0-h+d]$, and trace out the rightmost site. Under the
parity hypothesis and for homogeneous gates the set of window gates is the
same at every half-step, so a single $4^{d+1}\times4^{d+1}$ matrix
$\mathcal T$ suffices. There is one window gate for $d=1$, there are two for
$d=2$, and in general there are $\lceil(d+1)/2\rceil$. The readout acts on
the rightmost window site, which is $p_h$ by construction. The initial window
state is obtained by applying the layer-1 gates inside $[x_0-1,\,x_0+d]$ to
the product initial state and tracing out the rightmost site. Against full
density-matrix simulation this form agrees to $9.4\times10^{-16}$,
$3.9\times10^{-16}$ and $1.9\times10^{-16}$ at $d=0$, $1$ and $2$, while
applying the $d=1$ law to $d=0$ data, or the reverse, fails by
$2.4\times10^{-2}$. Figure~\ref{fig:hierprec} shows the agreement for each
displacement and Fig.~\ref{fig:hier} the means themselves.
Table~\ref{tab:ver2} lists the verification ensembles for the edge law and
the late twirl, and Table~\ref{tab:parity} the dependence on the sublattice
of the germ.

\begin{table}[t]\footnotesize
\caption{Verification ensembles for the edge law and the late twirl,
regenerated by \texttt{verify\_all.py}. Executed-schedule checks include
damping in every time slice. The controls in the lower block must fail.}
\label{tab:ver2}
\begin{ruledtabular}
\begin{tabular}{lc}
Ensemble & error\\ \hline
5 Haar gates $\times$ 4 germ vectors, $n=9$ & $1.3\times10^{-15}$\\
inhomogeneous, different $U$ at every step & $1.4\times10^{-16}$\\
$n$-independence, $n=8\ldots12$ at fixed geometry & $6.1\times10^{-16}$\\
unital depolarizing $p=0.5$ off ray & $3.1\times10^{-16}$\\
arbitrary $T_1$, $\gamma=0.3$, behind the ray & $1.4\times10^{-17}$\\
Thm.~\ref{thm:comoving}, $d=1,2,3$ (16, 64, 256 states) & $5.1\times10^{-16}$\\
executed schedule, late twirl, no noise & $2.8\times10^{-16}$\\
executed schedule, late twirl, $T_1$, $\gamma=0.15$ & $8.3\times10^{-17}$\\
late twirl, $T_1$ $\gamma=0.05$ / $0.2$ everywhere & $1.7\times10^{-16}$\\ \hline
$T_1$ $\gamma=0.05$ ahead of the ray & $6.0\times10^{-3}$\\
$T_1$ $\gamma=0.10$ ahead of the ray & $1.2\times10^{-2}$\\
$t=0$ twirl schedule, $T_1$ $\gamma=0.05$ every slice & $1.6\times10^{-2}$\\
$t=0$ twirl, best $4\times4$ model, $\gamma=0.02$ & $3.6\times10^{-3}$\\
untransposed law, transverse germ & $8.1\times10^{-3}$\\
readout displaced $+1$, Eq.~\eqref{eq:law} & $4.9\times10^{-2}$\\
readout displaced $+2$, Eq.~\eqref{eq:law} & $3.0\times10^{-2}$\\
\end{tabular}
\end{ruledtabular}
\end{table}

\begin{table}[t]
\caption{Dependence on the sublattice of the germ. The law holds when the
germ occupies the sublattice paired with the first brickwork layer and fails
otherwise, independently of the length of the chain. The size of the failure
depends on the gate, and we quote medians over 12 Haar-random gates with the
full range in brackets.}
\label{tab:parity}
\begin{ruledtabular}
\begin{tabular}{ccc}
$n$ & $x_0$ & error\\ \hline
9 & 7 (paired) & $1.4\times10^{-16}$\\
9 & 6 (unpaired) & $3.0\times10^{-2}$ $[5.0\times10^{-3},1.8\times10^{-1}]$\\
10 & 7 (paired) & $1.5\times10^{-16}$\\
10 & 8 (unpaired) & $3.0\times10^{-2}$ $[5.0\times10^{-3},1.8\times10^{-1}]$\\
\end{tabular}
\end{ruledtabular}
\end{table}

\section{The dual-unitary null at \texorpdfstring{$d=1$}{d=1}}
\label{app:dunull}

The null follows in one line from the co-moving frame. At $d=1$ the measured
qubit is the right output of a gate whose left input is a fresh
$\tfrac{\id}{2}$. For a dual-unitary gate the map from the right input to the
right output with a maximally mixed left input,
$\rho\mapsto\Tr_1[U(\tfrac{\id}{2}\otimes\rho)U^\dagger]$, is completely
depolarizing, so it sends any unnormalized state $\sigma$ of the co-moving
qubit to $\Tr(\sigma)\,\tfrac{\id}{2}$. Each outcome then occurs with
probability $\tfrac12$ whatever the past, and the record is exactly a
sequence of fair coins, $p(\bm m)=2^{-S}$. In the language of correlation
functions, the $d=1$ readouts lie on neither edge of the cone of the germ and
on neither edge of each other's cones, where dual-unitary circuits have no
correlations~\cite{BKP2019}. All means, all correlators and the
total-variation distance to the uniform distribution on $\{\pm1\}^S$ vanish
identically, and the computed distance at $S=4$ is $0.0000$. At $d=2$ the
means still vanish, while $\langle m_1m_{1+t}\rangle$ reproduces the $d=0$
values $0.6066$, $0.3679$ and $0.2232$ exactly, because consecutive readouts
are again connected by a light-cone edge. For a Haar-random gate with
dual-unitarity defect $1.91$ nothing vanishes at any $d$.

The practical consequence is that a $d=1$ rail measures the
dual-unitarity defect of the executed gate against an exact zero, with no
fitted baseline and no simulator, and Theorem~\ref{thm:comoving} supplies the
exact law for whatever nonzero signal appears. Coherent miscalibration of the
compiled three-CNOT template shows up directly. Unital noise outside the
strip does not (Appendix~\ref{app:comoving}).

\section{The germ-antisymmetrized witness}
\label{app:witness}

Record probabilities are linear in the initial state. Since
$\rho_0(\bm r)=\rho_0(0)+\tfrac12(\bm r\cdot\bm\sigma)_{x_0}\otimes
\bigotimes_{q\neq x_0}\tfrac{\id}{2}$, every record probability, for any
noisy circuit and any readout error, has the form
$p(\bm m|\bm r)=p(\bm m|0)+\bm r\cdot\bm g(\bm m)$. The antisymmetrized
record $\tfrac12[p(\bm m|\bm r)-p(\bm m|-\bm r)]=\bm r\cdot\bm g(\bm m)$
removes $p(\bm m|0)$ exactly, and with it every effect that does not depend
on the germ, including relaxation toward $|0\rangle$ and asymmetric readout
error. What remains is the linear response of the record to the germ. For a
dual-unitary gate at $d=1$ this response vanishes identically in the
noiseless circuit (Appendix~\ref{app:dunull}), and it remains zero when the
gates are dressed by single-qubit unitaries, which preserve dual-unitarity.
We verified to machine precision that it also remains zero under
depolarizing noise and amplitude damping on the rail window after every
gate, amplitude damping on the window after every readout, amplitude damping
on every qubit in every time slice, and asymmetric readout assignment errors.

\emph{Bound on the executed gate.} Candidate gates are parametrized as
\begin{align*}
U(\delta_a,\delta_b)={}&(u_+\otimes u_-)\,e^{-i[(\pi/4-\delta_a)XX+
(\pi/4-\delta_b)YY+JZZ]}\\
&\times(v_-\otimes v_+),
\end{align*}
with the dressing of the deployed gate. The rail noise is calibrated by
fitting the noisy law, with a depolarizing strength and amplitude damping
after each gate, amplitude damping after each readout and asymmetric
assignment errors, to the $d=0$ dual-unitary data of the same job. The fit
gives a depolarizing strength of $0.012$, damping of $0.000$ after the gate
and $0.039$ after the readout, and assignment errors of $0.0045$ and
$0.0035$. For each candidate the $d=1$ record is computed from
Theorem~\ref{thm:comoving} with this noise, and
$\chi^2(\delta)=\sum_h[(W_h^{\rm obs}-W_h(\delta))/\sigma_h]^2$ is compared
with its minimum. The $95\%$ interval, $\chi^2-\chi^2_{\min}\le3.84$ on a
grid of spacing $5\times10^{-4}$\,rad, is $\delta\in[-0.0110,+0.0085]$\,rad
for a common displacement $\delta_a=\delta_b=\delta$ and
$\delta\in[-0.0055,+0.0100]$\,rad for a differential one,
$\delta_a=-\delta_b=\delta$. The dual-unitarity defect at the edges of these
intervals is $0.042$ and $0.039$. At $\delta=0.05$\,rad the predicted witness
is $(-0.032,-0.036,0.011,-0.022)$ for the common direction and
$(0.055,0.027,0.026,0.018)$ for the differential one, several times the
statistical error of $0.0055$.

\section{Cloud platform and methods}
\label{app:methods}

All hardware runs used processors of IBM Quantum, accessed through the cloud
with Qiskit Runtime. Since the devices were not under our control and their
characteristics change over time, we document here the processors, the dates,
the job identifiers, the calibration data reported by the provider at the
time of each run and the software settings, following the guidelines of the
Physical Review journals for cloud quantum computing demonstrations. The
calibration records and job identifiers are stored with the data.

\emph{Processors.} \textit{ibm\_fez}, \textit{ibm\_marrakesh} and
\textit{ibm\_kingston} are 156-qubit superconducting processors of IBM's
Heron~r2 generation, with transmon qubits on a heavy-hexagonal lattice,
tunable couplers, the native gate set CZ, $R_z$, $\sqrt X$ and $X$, and
support for mid-circuit measurement and reset. The configuration records of
the provider give the processor family Heron, revision~2, for all three, and
its calibration records give the backend versions 1.3.37 for
\textit{ibm\_fez}, 1.0.21 for \textit{ibm\_marrakesh} and 1.0.0 for
\textit{ibm\_kingston} at the times of the runs. The device clock had the
period $\mathrm{dt}=4$\,ns on all three. On the chain of the two jobs of
September on \textit{ibm\_fez} a CZ gate took $68$\,ns, except on the bond
between qubits $146$ and $147$, where it took $88$\,ns, and a readout took
$1.66\,\mu$s. Every circuit of this work
was compiled to the native gate set by the Qiskit preset pass manager at
optimization level~1, with the chain of physical qubits pinned as the initial
layout, and was executed with the Sampler~V2 primitive of Qiskit Runtime in
job mode. The job records of the provider name the client software, Qiskit
2.5.1 with qiskit-ibm-runtime 0.48.0 for the runs of August 2026
and Qiskit 2.4.2 with qiskit-ibm-runtime 0.47.0 for the jobs of September.
Table~\ref{tab:platform} lists the runs.

\begin{table*}[t]\footnotesize
\caption{The hardware runs analysed in this work, all in 2026, with
submission times in UTC. The runbook of August did not store job
identifiers, and those given for August were identified afterwards as
described in the text.}
\label{tab:platform}
\begin{ruledtabular}
\begin{tabular}{lllllr}
run & data file & processor & job & submitted & qubits, shots\\ \hline
\multicolumn{6}{l}{\emph{Recorded rails (Sec.~\ref{sec:rails} and Table~\ref{tab:rails})}}\\
R1 & \texttt{r12\_clean} & \textit{ibm\_fez} & \texttt{d9ph073bvhrs73a2k7d0} & 5 Aug 10:27 & 12, $2\times10^4$\\
R2 & \texttt{r12\_v2} & \textit{ibm\_fez} & \texttt{d9phbjd7skus73ckvhqg} & 5 Aug 10:51 & 12, $2\times10^4$\\
R3 & \texttt{results\_n12} & \textit{ibm\_fez} & \texttt{d9pentu28h6s739r7fvg} & 5 Aug 07:53 & 12, $2\times10^4$\\
R4 & \texttt{r20\_clean} & \textit{ibm\_fez} & \texttt{d9ph0espcc1s73a7lhag} & 5 Aug 10:28 & 20, $2\times10^4$\\
R5 & \texttt{r20\_generic} & \textit{ibm\_fez} & \texttt{d9pj5rm28h6s739rd330} & 5 Aug 12:56 & 20, $2\times10^4$\\
R6 & \texttt{inj\_0\_0} & \textit{ibm\_fez} & \texttt{d9phfmcpcc1s73a7m6r0} & 5 Aug 11:00 & 20, $2\times10^4$\\
R7 & \texttt{r20\_haar2} & \textit{ibm\_fez} & \texttt{d9pj9jd7skus73cl21i0} & 5 Aug 13:04 & 20, $2\times10^4$\\
R8 & \texttt{results\_n20} & \textit{ibm\_fez} & \texttt{d9peoh57skus73cks790} & 5 Aug 07:54 & 20, $2\times10^4$\\
R9 & \texttt{r50\_clean} & \textit{ibm\_fez} & \texttt{d9pgvbspcc1s73a7lfhg} & 5 Aug 10:25 & 50, $2\times10^4$\\
R10 & \texttt{results\_n50} & \textit{ibm\_fez} & \texttt{d9pesud7skus73cksd10} & 5 Aug 08:04 & 50, $2\times10^4$\\
R11 & \texttt{r\_big} & \textit{ibm\_fez} & \texttt{d9ph79bbvhrs73a2kkdg} & 5 Aug 10:42 & 54, $2\times10^4$\\
R12 & \texttt{r\_100} & \textit{ibm\_fez} & \texttt{d9pk84bbvhrs73a2off0} & 5 Aug 14:09 & 54, $2\times10^4$\\
\multicolumn{6}{l}{\emph{Pauli noise injected off the ray (Sec.~\ref{sec:rails}) and qubit ordering (this appendix)}}\\
$p=0.1$ & \texttt{inj\_0\_1} & \textit{ibm\_fez} & \texttt{d9phhpm28h6s739rb0h0} & 5 Aug 11:05 & 20, $2\times10^4$\\
$p=0.2$ & \texttt{inj\_0\_2} & \textit{ibm\_fez} & \texttt{d9phi4628h6s739rb100} & 5 Aug 11:05 & 20, $2\times10^4$\\
$p=0.3$ & \texttt{inj\_0\_3} & \textit{ibm\_fez} & \texttt{d9phicspcc1s73a7m9tg} & 5 Aug 11:06 & 20, $2\times10^4$\\
$p=0.5$ & \texttt{inj\_0\_5} & \textit{ibm\_fez} & \texttt{d9phikd7skus73ckvpl0} & 5 Aug 11:06 & 20, $2\times10^4$\\
mirrored Haar gate & \texttt{r20\_haar} & \textit{ibm\_fez} & \texttt{d9pj7dspcc1s73a7odhg} & 5 Aug 12:59 & 20, $2\times10^4$\\
\multicolumn{6}{l}{\emph{Planted faults (Secs.~\ref{sec:rails} and \ref{sec:diagnosis}, Appendix~\ref{app:markov})}}\\
step 3, no control & \texttt{r20\_plant3\_marrakesh} & \textit{ibm\_marrakesh} & \texttt{d9pl28l7skus73cl4r9g} & 5 Aug 15:05 & 20, $2\times10^4$\\
control & \texttt{plant\_control} & \textit{ibm\_marrakesh} & \texttt{d9pl5ku28h6s739rg6p0} & 5 Aug 15:12 & 20, $2\times10^4$\\
step 3 & \texttt{plant\_step3\_v3} & \textit{ibm\_marrakesh} & \texttt{d9plbgjbvhrs73a2qae0} & 5 Aug 15:24 & 20, $2\times10^4$\\
step 3, repeated & \texttt{plant\_step3\_v4} & \textit{ibm\_marrakesh} & \texttt{d9plcnl7skus73cl5a8g} & 5 Aug 15:27 & 20, $2\times10^4$\\
step 2 & \texttt{plant\_step2} & \textit{ibm\_marrakesh} & \texttt{d9plkpe28h6s739rgp9g} & 5 Aug 15:44 & 20, $2\times10^4$\\
step 2, paired & \texttt{paired\_step2} & \textit{ibm\_marrakesh} & \texttt{d9plmukpcc1s73a7s3pg} & 5 Aug 15:49 & 20, $4\times10^4+4\times10^4$\\
step 2, paired & \texttt{paired\_step2\_fez} & \textit{ibm\_fez} & \texttt{d9plo757skus73cl5of0} & 5 Aug 15:51 & 20, $4\times10^4+4\times10^4$\\
\multicolumn{6}{l}{\emph{Stuck readout, one chain (Secs.~\ref{sec:rails} and \ref{sec:diagnosis})}}\\
stuck readout & \texttt{r50\_S4} & \textit{ibm\_marrakesh} & \texttt{d9pfd257skus73cksv70} & 5 Aug 08:38 & 50, $2\times10^4$\\
stuck readout & \texttt{r50\_nodd} & \textit{ibm\_marrakesh} & \texttt{d9pfgv628h6s739r8c90} & 5 Aug 08:46 & 50, $2\times10^4$\\
stuck readout & \texttt{r50\_b} & \textit{ibm\_marrakesh} & \texttt{d9pgj8rbvhrs73a2jh90} & 5 Aug 10:00 & 50, $2\times10^4$\\
\multicolumn{6}{l}{\emph{Interleaved and registered demonstrations (Secs.~\ref{sec:interleaved} to \ref{sec:memory}, Appendices~\ref{app:interleaved} and \ref{app:memreg})}}\\
interleaved & \texttt{fez\_lightcone} & \textit{ibm\_fez} & \texttt{daofjddr85ps73ff7970} & 21 Sep 09:32 & 20, $262\,144$\\
registered & \texttt{memory\_fez} & \textit{ibm\_fez} & \texttt{daop43opqrnc739aancg} & 21 Sep 20:22 & 20, $589\,824$\\
repetition & \texttt{memory\_marrakesh} & \textit{ibm\_marrakesh} & \texttt{daop93opqrnc739ab0og} & 21 Sep 20:33 & 20, $589\,824$\\
repetition & \texttt{memory\_kingston} & \textit{ibm\_kingston} & \texttt{daopb1n8gn2s739o77g0} & 21 Sep 20:37 & 20, $589\,824$\\
\end{tabular}
\end{ruledtabular}
\end{table*}

\emph{Job identifiers.} The runbook used in August 2026 stored the records
but not the job identifiers. All twenty-seven runs of August analysed here
were identified afterwards in the records of the provider. For each of them
the outcome histogram of every circuit of the job, of $1250$ or $2500$
shots, coincides with the corresponding block of the stored records.
Whatever the outcome probabilities, the chance that a run reproduces a given
histogram exactly is largest when they equal the observed frequencies, and
even then the chance that another run reproduces all histograms of one of
these jobs is below $10^{-250}$. The original times of the data files, kept
with the data, agree, since every one of these jobs finished before the
runbook wrote its file, less than $18$ minutes earlier. The four jobs of
September 2026 recorded their identifiers at submission, and for them too the
histogram of every circuit returned by the provider coincides with the
corresponding block of the stored records. Every job started to run within
$90$\,s of submission, except R12, which waited $63$ minutes in the queue,
the paired run on \textit{ibm\_fez}, which waited $11$ minutes, and R11, R6
and R2, which waited $9$, $4$ and $3$ minutes.

\emph{Calibration data.} For every job we retrieved from the
provider the calibration record in force when the job started to run, for
all qubits and bonds of its chain, and these records are stored with the
data. Table~\ref{tab:calsum} summarizes those of the runs analysed here, and
Table~\ref{tab:calib} lists in full the records for the two jobs of September
on
\textit{ibm\_fez}, last updated at 09:03 and 19:08~UTC on the day of the
jobs. The records contain no qubit frequencies. In the provider's history a
record can take effect minutes after its last update, at least $13$ minutes in
one case here. The records retrieved show six runs, R1, R4, \texttt{r50\_S4}
and three planted-fault runs, that started under a record although a newer
one had been last updated $71$ to $785$\,s before. For four of the six that
newer record is stored, with later runs, for the whole chain, and it differs
from the one in force at most in the single-qubit gate errors, which
Table~\ref{tab:calsum} does not list. For the four jobs of
September the runbook had also stored, at submission, the readout assignment
error and the relaxation time $T_1$ of every rail qubit, and the records
agree with these values exactly. The chain of the interleaved job was
selected with cuts of $3\%$ on the readout error and $60\,\mu$s on $T_1$
over the rail region, chain positions $13$ to $18$, and ranked by the summed
readout and two-qubit gate errors there, and the registered job used the
same qubits in reverse order. The chains of August were selected at
submission, against the calibration then published, with cuts of $5\%$ on
the readout error and $40\,\mu$s on $T_1$ over the $S+1$ rail sites, and the
device excluded $16$ qubits by those cuts on the day of the 50-qubit runs. In
the records in force at execution every rail site of the runs of August
passes these cuts, except qubit $65$, the last rail site of R11 and R12 and
the germ of R10, at readout errors of $5.4\%$ to $5.7\%$, qubit $43$, a rail
site of R10, at $8.1\%$, and qubit $11$, the germ of R3, at $7.8\%$.

\begin{table*}[t]\footnotesize
\caption{Calibration record of the provider for the twenty qubits of the
chain of the two jobs on \textit{ibm\_fez}, in force when the first job ran
(last update 09:03~UTC) and when the second ran (19:08~UTC) on 21 September
2026, within two seconds of their submission. Pairs of values refer to the first and the second job, and
$x$ gives the chain position in each. The readout assignment error $q$ was
the same in both records. $\epsilon_{\sqrt X}$ is the error of the $\sqrt X$
gate and $\epsilon_{\rm CZ}$ that of the CZ gate on the bond to the qubit of
the next row, which took $68$\,ns, or $88$\,ns where marked by $\dagger$. In
both jobs the germ sits at $x=17$ and the edge rail at $x=13$ to $16$.}
\label{tab:calib}
\begin{ruledtabular}
\begin{tabular}{rcccccc}
qubit & $x$ & $T_1$ ($\mu$s) & $T_2$ ($\mu$s) & $q$ (\%) & $\epsilon_{\sqrt X}$ ($10^{-4}$) & $\epsilon_{\rm CZ}$ ($10^{-3}$)\\ \hline
98 & 0, 19 & 136\,/\,124 & 24\,/\,22 & 1.25 & 3.4\,/\,2.4 & 3.0\,/\,2.7\\
111 & 1, 18 & 182\,/\,198 & 113\,/\,123 & 0.63 & 2.5\,/\,2.6 & 2.6\,/\,2.3\\
110 & 2, 17 & 121\,/\,144 & 116\,/\,116 & 0.29 & 2.9\,/\,2.0 & 2.3\,/\,2.2\\
109 & 3, 16 & 166\,/\,162 & 69\,/\,71 & 0.61 & 3.6\,/\,3.4 & 2.4\,/\,2.9\\
118 & 4, 15 & 141\,/\,82 & 58\,/\,48 & 0.54 & 2.1\,/\,4.0 & 6.3\,/\,2.6\\
129 & 5, 14 & 132\,/\,103 & 109\,/\,91 & 0.85 & 2.4\,/\,2.1 & 8.1\,/\,2.9\\
130 & 6, 13 & 80\,/\,152 & 110\,/\,162 & 0.51 & 4.2\,/\,4.3 & 2.2\,/\,2.1\\
131 & 7, 12 & 206\,/\,159 & 245\,/\,160 & 7.03 & 1.3\,/\,1.7 & 2.3\,/\,2.4\\
138 & 8, 11 & 82\,/\,81 & 100\,/\,86 & 0.61 & 4.0\,/\,3.1 & 2.4\,/\,4.6\\
151 & 9, 10 & 87\,/\,71 & 116\,/\,101 & 1.05 & 1.8\,/\,3.0 & 7.6\,/\,7.0\\
150 & 10, 9 & 83\,/\,141 & 5.4\,/\,5.3 & 0.49 & 7.2\,/\,7.9 & 27.2\,/\,45.3\\
149 & 11, 8 & 49\,/\,67 & 5.1\,/\,4.8 & 5.10 & 162\,/\,162 & 26.4\,/\,26.4\\
148 & 12, 7 & 108\,/\,126 & 145\,/\,137 & 0.83 & 3.9\,/\,3.1 & 3.9\,/\,2.5\\
147 & 13, 6 & 102\,/\,177 & 105\,/\,139 & 0.42 & 5.5\,/\,3.0 & 3.6\,/\,2.2$^\dagger$\\
146 & 14, 5 & 112\,/\,106 & 44\,/\,16 & 0.37 & 1.7\,/\,2.2 & 1.6\,/\,1.8\\
145 & 15, 4 & 155\,/\,196 & 124\,/\,135 & 1.39 & 1.9\,/\,1.6 & 1.8\,/\,1.7\\
144 & 16, 3 & 111\,/\,139 & 158\,/\,193 & 0.68 & 2.9\,/\,2.8 & 2.1\,/\,2.1\\
143 & 17, 2 & 131\,/\,158 & 154\,/\,156 & 0.59 & 2.0\,/\,2.0 & 1.9\,/\,1.8\\
142 & 18, 1 & 132\,/\,200 & 187\,/\,234 & 0.37 & 2.7\,/\,4.4 & 1.8\,/\,1.9\\
141 & 19, 0 & 120\,/\,127 & 83\,/\,85 & 0.90 & 1.7\,/\,2.0 & \\
\hline
median & & 120\,/\,140 & 109\,/\,108 & 0.62 & 2.8\,/\,2.9 & 2.4\,/\,2.4\\
\end{tabular}
\end{ruledtabular}
\end{table*}

\emph{Poorly calibrated qubits.} The provider's records show three poorly
calibrated qubits on the chain of the two jobs of September on
\textit{ibm\_fez}. Qubit $149$ had $T_1$ of $49$ and
$67\,\mu$s, $T_2\approx5\,\mu$s, a readout assignment error of $5.1\%$ and a
$\sqrt X$ error of $1.6\%$, qubit $150$ had $T_2\approx5\,\mu$s, and qubit
$131$ a readout assignment error of $7.0\%$. The CZ gates between $148$ and
$149$ and between $149$ and $150$ had errors of $2.6\%$ to $4.5\%$. These
qubits lie at chain positions $7$ to $12$, ahead of the last rail site at
position $13$, and none of them contributes a record bit. In every
configuration with the late twirl they are therefore exactly outside the
support of the record law, by Theorem~\ref{thm:noise}(iii), and this holds
for the displaced records as well, since only the line $L_0$ is fed from
outside the cone. Only in the configurations of the first job with the twirl
at $t=0$ can $149$ and $150$, at positions $11$ and $10$, reach the record at
all, through non-unital noise passed on by two or three gates to the site
that joins the ray at $h=4$. These are the configurations built to expose
such noise (Sec.~\ref{sec:idle}). Qubit $98$, with $T_2=22\,\mu$s in the
second job, enters only its records at $d=2$, as a maximally mixed input of
the first gate of $L_1$, on which dephasing has no effect.

The chains of August pass through poorly calibrated qubits as well. The
50-qubit chain of the stuck-readout runs contains, at positions $8$ to $16$,
qubits with readout errors of up to $43\%$ and two bonds with a reported CZ
error of $1$, and in the 50-qubit chain of R9 qubit $72$, at position $28$,
had a readout error of $32\%$, $T_1=6\,\mu$s and no reported $T_2$, with a
CZ error of $1$ on both of its bonds. R12 contains a qubit with a readout
error of $48\%$ at position $33$, and the chain shared by R5, R7 and the
mirrored-gate run one with $13\%$ at position $7$. All of them lie at least
eight sites ahead of the last rail site, beyond the reach of the record. In
these rails the background is randomized only at $t=0$, so relaxation on the
$S-1$ sites ahead of the last rail site, three for $S=4$, can still reach the
record at the last step (Theorem~\ref{thm:noise}(ii)). On those sites the
records list $T_1$ below $40\,\mu$s for qubit $64$ of R11 and R12, at
$32\,\mu$s, and for qubit $12$ of R8, at $33\,\mu$s, and these rails
nevertheless fit the step-dependent chain at $\chi^2/\mathrm{dof}=1.75$,
$0.27$ and $0.61$ (Table~\ref{tab:rails}). Qubit $63$ of R11, two sites
ahead of its last rail site, had a readout error of $10\%$, which does not
enter the record, since that qubit is not read. Nor do the records single
out R6 and R9, the two rails above $3$ in Table~\ref{tab:rails}, on their
rails or within reach of them, and R6 ran on the chain of R4, which fits at
$1.27$.

\begin{table*}[t]\footnotesize
\caption{Calibration records of the provider for every run analysed in
this work, in force when the run started. The columns \emph{ran} and
\emph{calibrated} give the time the run started and that of the last update
of the record, in UTC. Each record covers all qubits
and bonds of the chain of $n$ qubits, and the entries give the median over
the chain and, in brackets, the smallest $T_1$ and $T_2$ and the largest
readout assignment error $q$ and CZ error $\epsilon_{\rm CZ}$. A CZ error of
$1000\times10^{-3}$ is the value $1$ reported for two bonds in each of the
50-qubit chains of the stuck-readout runs and of R9, and no $T_2$ is reported
for qubit $72$ of the chain of R9. The last column gives the largest $q$ (in \%) and the smallest $T_1$
(in $\mu$s) over the $S+1$ rail sites of the runs of August and over the rail
region, positions $13$ to $18$, of the jobs of September.}
\label{tab:calsum}
\begin{ruledtabular}
\setlength{\tabcolsep}{4pt}
\begin{tabular}{llrllccccc}
run & processor & $n$ & ran & calibrated & $T_1$ ($\mu$s) & $T_2$ ($\mu$s) & $q$ (\%) & $\epsilon_{\rm CZ}$ ($10^{-3}$) & rail $q$, $T_1$\\ \hline
\multicolumn{10}{l}{\emph{5 August 2026}}\\
R1 & \textit{ibm\_fez} & 12 & 10:27 & 09:11 & 112\,[52] & 67\,[15] & 1.38\,[8.2] & 2.6\,[3.3] & 3.30, 52\\
R2 & \textit{ibm\_fez} & 12 & 10:54 & 10:26 & 150\,[52] & 94\,[18] & 0.99\,[4.9] & 2.4\,[8.4] & 4.91, 52\\
R3 & \textit{ibm\_fez} & 12 & 07:53 & 07:38 & 153\,[53] & 86\,[21] & 0.98\,[7.8] & 2.7\,[10.3] & 7.85, 91\\
R4 & \textit{ibm\_fez} & 20 & 10:28 & 09:11 & 110\,[52] & 90\,[18] & 0.70\,[5.0] & 2.9\,[5.5] & 4.91, 52\\
R5 & \textit{ibm\_fez} & 20 & 12:56 & 12:38 & 108\,[67] & 109\,[24] & 1.06\,[13.0] & 2.8\,[8.7] & 2.10, 86\\
R6 & \textit{ibm\_fez} & 20 & 11:04 & 10:26 & 110\,[52] & 90\,[18] & 0.70\,[5.0] & 2.9\,[5.5] & 4.91, 52\\
R7 & \textit{ibm\_fez} & 20 & 13:04 & 12:38 & 108\,[67] & 109\,[24] & 1.06\,[13.0] & 2.8\,[8.7] & 2.10, 86\\
R8 & \textit{ibm\_fez} & 20 & 07:54 & 07:38 & 142\,[33] & 75\,[21] & 1.04\,[7.8] & 2.7\,[10.3] & 1.81, 54\\
R9 & \textit{ibm\_fez} & 50 & 10:25 & 09:11 & 121\,[6.2] & 102\,[4.7] & 0.84\,[32.1] & 2.8\,[1000] & 1.73, 48\\
R10 & \textit{ibm\_fez} & 50 & 08:04 & 07:38 & 134\,[33] & 92\,[15] & 0.99\,[9.2] & 2.7\,[39.7] & 8.13, 85\\
R11 & \textit{ibm\_fez} & 54 & 10:51 & 10:26 & 121\,[29] & 99\,[4.7] & 0.98\,[10.0] & 2.8\,[39.7] & 5.42, 74\\
R12 & \textit{ibm\_fez} & 54 & 15:12 & 13:28 & 121\,[29] & 99\,[4.7] & 0.96\,[48.4] & 2.8\,[66.0] & 5.65, 74\\
injection, $p=0.1$ & \textit{ibm\_fez} & 20 & 11:05 & 10:26 & 110\,[52] & 90\,[18] & 0.70\,[5.0] & 2.9\,[5.5] & 4.91, 52\\
injection, $p=0.2$ & \textit{ibm\_fez} & 20 & 11:05 & 10:26 & 110\,[52] & 90\,[18] & 0.70\,[5.0] & 2.9\,[5.5] & 4.91, 52\\
injection, $p=0.3$ & \textit{ibm\_fez} & 20 & 11:06 & 10:26 & 110\,[52] & 90\,[18] & 0.70\,[5.0] & 2.9\,[5.5] & 4.91, 52\\
injection, $p=0.5$ & \textit{ibm\_fez} & 20 & 11:06 & 10:26 & 110\,[52] & 90\,[18] & 0.70\,[5.0] & 2.9\,[5.5] & 4.91, 52\\
mirrored Haar gate & \textit{ibm\_fez} & 20 & 12:59 & 12:38 & 108\,[67] & 109\,[24] & 1.06\,[13.0] & 2.8\,[8.7] & 2.10, 86\\
plant at step 3, no control & \textit{ibm\_marrakesh} & 20 & 15:06 & 12:04 & 186\,[44] & 99\,[13] & 0.77\,[3.9] & 2.0\,[8.8] & 0.85, 84\\
plant control & \textit{ibm\_marrakesh} & 20 & 15:12 & 12:04 & 167\,[44] & 108\,[13] & 1.02\,[7.0] & 2.2\,[8.8] & 3.08, 82\\
plant at step 3 & \textit{ibm\_marrakesh} & 20 & 15:24 & 15:04 & 167\,[44] & 108\,[13] & 1.02\,[7.0] & 2.2\,[8.8] & 3.08, 82\\
plant at step 3, repeated & \textit{ibm\_marrakesh} & 20 & 15:27 & 15:04 & 167\,[44] & 108\,[13] & 1.02\,[7.0] & 2.2\,[8.8] & 3.08, 82\\
plant at step 2 & \textit{ibm\_marrakesh} & 20 & 15:44 & 15:04 & 167\,[44] & 108\,[13] & 1.02\,[7.0] & 2.2\,[8.8] & 3.08, 82\\
plant at step 2, paired & \textit{ibm\_marrakesh} & 20 & 15:49 & 15:31 & 167\,[44] & 108\,[13] & 1.02\,[7.0] & 2.2\,[8.8] & 3.08, 82\\
plant at step 2, paired & \textit{ibm\_fez} & 20 & 16:03 & 15:40 & 111\,[54] & 93\,[18] & 1.10\,[5.6] & 2.9\,[6.4] & 4.08, 82\\
stuck readout, \texttt{r50\_S4} & \textit{ibm\_marrakesh} & 50 & 08:39 & 07:56 & 161\,[7.0] & 78\,[7.0] & 1.03\,[43.0] & 3.2\,[1000] & 2.58, 141\\
stuck readout, \texttt{r50\_nodd} & \textit{ibm\_marrakesh} & 50 & 08:46 & 08:27 & 161\,[7.0] & 78\,[7.0] & 1.03\,[43.0] & 3.2\,[1000] & 2.58, 141\\
stuck readout, \texttt{r50\_b} & \textit{ibm\_marrakesh} & 50 & 10:00 & 09:32 & 161\,[7.0] & 78\,[7.0] & 1.03\,[43.0] & 3.2\,[1000] & 2.58, 141\\
\multicolumn{10}{l}{\emph{21 September 2026}}\\
interleaved & \textit{ibm\_fez} & 20 & 09:32 & 09:03 & 120\,[49] & 109\,[5.1] & 0.62\,[7.0] & 2.4\,[27.2] & 1.39, 102\\
registered & \textit{ibm\_fez} & 20 & 20:22 & 19:08 & 140\,[67] & 108\,[4.8] & 0.62\,[7.0] & 2.4\,[45.3] & 0.85, 82\\
repetition & \textit{ibm\_marrakesh} & 20 & 20:33 & 01:28 & 196\,[68] & 113\,[31] & 0.69\,[9.9] & 1.9\,[12.8] & 0.81, 102\\
repetition & \textit{ibm\_kingston} & 20 & 20:37 & 20:15 & 262\,[125] & 167\,[30] & 0.60\,[3.8] & 2.0\,[25.1] & 0.61, 208\\
\end{tabular}
\end{ruledtabular}
\end{table*}

\emph{Error mitigation, decoupling and twirling.} No error mitigation of
any kind was applied in any run, and no measurement mitigation. In the four
jobs of September, dynamical decoupling, gate twirling and measurement
twirling were switched off explicitly in the Sampler options, as the job
records confirm. The runbook that recorded the runs of August 2026 could pad
idle qubits with a CPMG sequence of two $X$ gates, and the circuits held by
the provider show where it did. Among the twenty-seven runs of August analysed
here only \texttt{r50\_S4} carries the sequence, with delays and $X$ gates
on idle qubits. The other twenty-six, among them every recorded rail,
\texttt{r50\_nodd} and every planted-fault run, contain neither delays nor
$X$ gates. The circuits of the
September jobs contain no decoupling sequence, and their only delays are the
injected idle periods of the interleaved job, one on each chain qubit in
each of the $32$ circuits with injection. A decoupling sequence consists of
single-qubit unitaries. By Theorem~\ref{thm:noise} it cannot
change the record when it acts off the ray, and on the rail it is absorbed
into the step-dependent kernel, so the Markov-structure test of
Sec.~\ref{sec:rails} does not depend on it.

\emph{Shots and background.} The recorded rails used $2\times10^4$ shots
each, split over $16$ random patterns of the background randomization. The
current runbook prepares the background by a Hadamard gate and a mid-circuit
measurement on every background site, which is exact in each shot. The
dataset files do not record the preparation method, and the earliest rails
were taken with a runbook that randomized the background with random $X$
patterns. The Markov test is applied to all of them in the same way. In the
interleaved jobs the late twirl was realized by enumerating the $4^S=256$
Pauli assignments as a parameter sweep with equal shots, and the injected
idle period was a \texttt{delay} instruction of $12\,496$ samples of the
device clock ($\mathrm{dt}=4$\,ns), that is $49.98\,\mu$s.

\emph{Chain selection.} Chains of qubits were found by randomized
depth-first search with a Warnsdorff heuristic on the coupling map of the
device. Calibration quality was required only on the rail sites, since noise
behind the ray is irrelevant by Theorem~\ref{thm:noise}(i), and the
remaining qubits were left unconstrained. This is what made a 50-qubit chain
schedulable on a device on which $16$ qubits failed the cuts.

\emph{Transpilation.} The two-qubit gate count of every transpiled circuit
was compared with the analytic value for the intended layout, and jobs were
refused whenever the transpiler inserted SWAP gates, since a rerouted rail
voids the exact reference.

\emph{Qubit ordering.} Qiskit orders qubits little-endian, so a raw $U(4)$
appended to wires $(a,a+1)$ realizes $SUS$, the mirror image of the matrix
written with wire $a$ as the first tensor factor. The mirror is not harmless.
For the deployed dressing angles $\|\mathcal M-\mathcal M_{\rm mirror}\|
=2.35$, an $8\sigma$ systematic at $2\times10^4$ shots. In the recorded
dual-unitary rails the gate was compiled from an explicit three-CNOT
template on named wires, and rebuilding that template in exact arithmetic
gives $\|T-U\|=0$ and $\|T-SUS\|=2.69$ for the compilation flag used
(\texttt{endian\_resolve.py}). The Haar-random rails discriminate between
the two hypotheses by a factor above $20$ in $\chi^2$ and confirm the
assignment independently. In the interleaved demonstrations every gate was
appended as \texttt{unitary(U,[a+1,a])}, and the convention was checked
against the matrix before submission. Every dataset written by the current
runbook records the convention.

\emph{Statistics.} The primary test is the Markov structure of
Corollary~\ref{cor:chain}, fitted to the full distribution of $2^S$ outcomes
(Appendix~\ref{app:markov}). It assumes nothing about the gate or the form of
the noise. Kernel shifts are estimated from conditional means of the
recorded bits, whose readout-induced bias is common to configurations
measured on the same qubits and cancels in paired differences. Standard
errors are analytic and agree with bootstrap estimates to within a few
percent.

\emph{Fault localization.} Because $\langle m_1m_{1+k}\rangle$ involves only
the gates of steps $2,\dots,k+1$, the first lag that deviates points to one
gate, and shorter lags are provably unaffected. A uniform noise floor
deviates from the first lag onward, and it is distinguished from a single
fault by the absence of a clean prefix. On chains whose baseline error
already breaks the first lag, this absolute localization is uninformative. We
then calibrate $(\lambda,q)$ on the causal prefix and predict the remainder,
or take the difference of two runs on the same physical chain.

\section{The interleaved demonstration}
\label{app:interleaved}

\emph{Chain and calibration.} We used twenty physical qubits of
\textit{ibm\_fez} in the order 98, 111, 110, 109, 118, 129, 130, 131, 138,
151, 150, 149, 148, 147, 146, 145, 144, 143, 142, 141, with the germ at chain
position $x_0=17$, physical qubit $143$. The rail region, chain positions
$13$ to $18$, consists of physical qubits $147$ to $142$, whose reported
readout assignment errors at submission were $0.37\%$ to $1.39\%$ and whose
$T_1$ times were $102$ to $155\,\mu$s (Table~\ref{tab:calib}). The chain was selected by a randomized
search that required readout errors of at most $3\%$ and $T_1$ of at least
$60\,\mu$s on the rail region, and it was ranked by the summed readout and
two-qubit gate errors there.

\emph{Configurations.} The eight configurations comprise the dual-unitary
gate at $d=0$ with either twirl schedule, the Haar-random gate at $d=0$ with
either schedule, both with and without the injected idle, and both gates at
$d=1$ with the late twirl. Every configuration was run with the germ at
$+\bm r$ and $-\bm r$, $\bm r=(\sin0.3\pi,0,\cos0.3\pi)$, prepared by
$R_y(0.3\pi)$ and $R_y(1.3\pi)$.

\emph{Circuits.} Background sites are prepared by a Hadamard gate and a
mid-circuit measurement, on every non-germ site for the twirl at $t=0$ and on
every site outside the rail for the late twirl. The late twirl is
$R_x(\pi a)R_z(\pi b)$ on site $x_0-h$ immediately before layer $h$, with all
$4^S=256$ assignments of $(a_h,b_h)\in\{0,1\}^2$ enumerated as a parameter
sweep with $64$ shots each, which realizes the uniform Pauli average exactly.
The injected idle is a delay of $49.98\,\mu$s, an integer multiple of $16$
samples, on every chain qubit after the background preparation. Gates were
compiled by the transpiler at optimization level $1$ with the chain pinned as
the initial layout.

\emph{Checks before submission.} The emitted gate was compared with its
matrix, every circuit was simulated noiselessly at $n=8$ and compared with
the exact law, and every transpiled circuit was checked for the absence of
SWAP gates, an unpermuted final layout, two-qubit gates only between
neighbouring chain qubits, each record bit measured on the intended physical
qubit, and intact sweep parameters.

\emph{Execution.} Each of the $16$ circuits was split into $8$ blocks, with
$32$ of the $256$ twirl assignments per block for the late twirl and $2048$
shots per block for the twirl at $t=0$. The $128$ blocks were interleaved in
the order block, configuration, germ sign within one Sampler job
(\texttt{daofjddr85ps73ff7970}). No dynamical decoupling, no runtime
twirling and no error mitigation were used. The job returned $16\,384$ shots
per configuration and germ sign, $262\,144$ in total, with all blocks
complete.

\emph{Acceptance criteria.} The criteria, written down before submission,
are those stated in Sec.~\ref{sec:interleaved}, and all were met.
Table~\ref{tab:shift} lists the idle-induced kernel shifts.

\begin{table}[h]\footnotesize
\caption{Idle-induced kernel shifts, injected minus not injected, for the
Haar-random gate with both germ signs pooled.}
\label{tab:shift}
\begin{ruledtabular}
\begin{tabular}{lccccc}
twirl & $h$ & $\Delta t_h$ & $z$ & $\Delta s_h$ & $z$\\ \hline
$t=0$ & 2 & $-0.0164$ & $-2.5$ & $-0.0627$ & $-9.5$\\
$t=0$ & 3 & $+0.0093$ & $+1.3$ & $+0.0587$ & $+8.4$\\
$t=0$ & 4 & $+0.0080$ & $+1.2$ & $+0.0184$ & $+2.7$\\
late & 2 & $+0.0100$ & $+1.5$ & $+0.0068$ & $+1.0$\\
late & 3 & $-0.0046$ & $-0.7$ & $+0.0039$ & $+0.6$\\
late & 4 & $-0.0011$ & $-0.2$ & $-0.0062$ & $-0.9$\\
\end{tabular}
\end{ruledtabular}
\end{table}

\section{Memory in the first interleaved job}
\label{app:memfirst}

The order-two memory $\delta(h,a)$ of Eq.~\eqref{eq:memory} is estimated
from the recorded bits of each germ sign, and
$\Delta=\tfrac12[\delta_{+\bm r}-\delta_{-\bm r}]$. Standard errors are
bootstrap estimates over shots with $400$ resamples and agree with the
binomial estimate. Table~\ref{tab:memfirst} lists every late-twirl
configuration of the interleaved job. The prediction is the co-moving law of
Theorem~\ref{thm:comoving} without noise. Including the noise calibrated on
the $d=0$ dual-unitary record of the same job, a depolarizing strength of
$0.012$ per layer, relaxation of $0.039$ after each readout and assignment
errors of $0.0045$ and $0.0035$, moves the prediction for $\Delta(3,+)$ from
$0.052$ to $0.045$ and gives $\chi^2=3.1$ on four components, against $2.1$
without noise. The data do not distinguish between the two.

The germ-even part $\tfrac12[\delta_{+\bm r}+\delta_{-\bm r}]$ measures
something else. On the edge the recorded bit differs from the projected one
by the assignment error of its qubit, which makes the recorded chain a
hidden chain. Exact simulation of the Haar gate gives a germ-even memory of
$0.007$, $0.019$ and $0.037$ for assignment errors of $0.5\%$, $1.4\%$ and
$3\%$, and a germ-odd part of $6\times10^{-5}$, $4\times10^{-4}$ and
$1.6\times10^{-3}$. Pooled over the six $d=0$ configurations, the measured
germ-even memory at $h=3$ is $0.015\pm0.003$, the size expected from the
reported assignment error of $1.39\%$ of the qubit read at step~$2$. The
germ-odd part that this error induces is more than an order of magnitude
below the statistical error. The analysis was performed after the job and
was not among its acceptance criteria.

\begin{table}[h]\footnotesize
\caption{Germ-antisymmetrized order-two memory $\Delta(h,m_{h-1})$ in the
first interleaved job on \textit{ibm\_fez}, late twirl, $3.3\times10^4$
shots per configuration. The last two columns give $\chi^2$ on four
components against zero and against the parameter-free law. Standard errors
are $0.008$ throughout, and $0.010$ for the dual-unitary gate at $d=0$,
$h=3$. Regenerated by \texttt{memory\_hw.py}.}
\label{tab:memfirst}
\setlength{\tabcolsep}{4pt}
\begin{ruledtabular}
\begin{tabular}{lcccccc}
 & $(3,+)$ & $(3,-)$ & $(4,+)$ & $(4,-)$ & $\chi^2_0$ & $\chi^2_{\rm law}$\\ \hline
Haar, $d=1$ & $+0.051$ & $-0.005$ & $+0.022$ & $-0.004$ & $50.6$ & $2.1$\\
\quad law & $+0.052$ & $+0.005$ & $+0.016$ & $-0.004$ & & \\
DU, $d=1$ & $-0.006$ & $-0.004$ & $-0.008$ & $-0.010$ & $3.7$ & $3.7$\\
Haar, $d=0$ & $+0.003$ & $-0.004$ & $+0.000$ & $-0.010$ & $2.0$ & $2.0$\\
DU, $d=0$ & $-0.016$ & $-0.002$ & $-0.008$ & $-0.019$ & $7.7$ & $7.7$\\
\end{tabular}
\end{ruledtabular}
\end{table}

\begin{table*}[tbh]\footnotesize
\caption{Registered memory demonstration. Measured $\Delta(h,m_{h-1})$ with
binomial standard errors in units of the last digit, the law with the chain
noise, and the noiseless law. Bootstrap standard errors over shots agree with
the binomial ones to $3\%$, and an independent evaluation from the raw bits
reproduces every entry. Regenerated by \texttt{memory\_experiment.py
analyze}.}
\label{tab:memreg}
\begin{ruledtabular}
\begin{tabular}{llcccc}
record & & $(3,+)$ & $(3,-)$ & $(4,+)$ & $(4,-)$\\ \hline
DU, $d=0$ & measured & $+0.0065(70)$ & $-0.0001(71)$ & $-0.0009(58)$ & $+0.0003(61)$\\
DU, $d=1$ & measured & $-0.0028(55)$ & $-0.0041(55)$ & $-0.0079(55)$ & $+0.0044(55)$\\
DU, $d=2$ & measured & $+0.0001(52)$ & $-0.0065(58)$ & $+0.0065(55)$ & $-0.0093(57)$\\
Haar A, $d=0$ & measured & $+0.0150(57)$ & $-0.0127(58)$ & $+0.0040(55)$ & $+0.0007(56)$\\
Haar B, $d=0$ & measured & $+0.0002(55)$ & $+0.0025(55)$ & $-0.0055(55)$ & $+0.0012(55)$\\ \hline
Haar A, $d=1$ & measured & $+0.0404(55)$ & $-0.0030(56)$ & $+0.0216(55)$ & $-0.0077(55)$\\
& law, chain noise & $+0.0370$ & $+0.0050$ & $+0.0106$ & $-0.0023$\\
& law, noiseless & $+0.0522$ & $+0.0048$ & $+0.0160$ & $-0.0037$\\
Haar A, $d=2$ & measured & $-0.0217(57)$ & $-0.0184(54)$ & $+0.0095(56)$ & $-0.0124(55)$\\
& law, chain noise & $-0.0148$ & $-0.0192$ & $-0.0048$ & $-0.0055$\\
& law, noiseless & $-0.0230$ & $-0.0314$ & $-0.0080$ & $-0.0091$\\
Haar B, $d=1$ & measured & $-0.0736(75)$ & $+0.0911(73)$ & $-0.0033(55)$ & $+0.0054(55)$\\
& law, chain noise & $-0.0718$ & $+0.0741$ & $-0.0052$ & $+0.0088$\\
& law, noiseless & $-0.1197$ & $+0.1179$ & $-0.0084$ & $+0.0137$\\
Haar B, $d=2$ & measured & $+0.0103(56)$ & $-0.0111(57)$ & $-0.0353(55)$ & $+0.0985(56)$\\
& law, chain noise & $-0.0027$ & $-0.0158$ & $-0.0228$ & $+0.0888$\\
& law, noiseless & $-0.0075$ & $-0.0245$ & $-0.0367$ & $+0.1645$\\
\end{tabular}
\end{ruledtabular}
\end{table*}

\section{The registered memory demonstration}
\label{app:memreg}

\emph{Registration.} The configurations, gates, seeds, shot numbers,
observable, error estimate, noise model, thresholds and criteria K1 to K5
were written down in a protocol before the job was
submitted, together with the SHA-256 checksum of the analysis script
\texttt{memory\_experiment.py} (\texttt{06e1803b\ldots97a1ea}). The script
that analysed the data has that checksum. The protocol was not deposited
publicly before the job, so its date rests on the author's records, which
give 20:19~UTC on 21 September 2026, three minutes before the job was
created. It is available with the code. Before
submission we estimated its
power on synthetic device data whose noise lies outside the fitted family and
is harsher than that of the first job. A correct theory meets K1 to K4
jointly in $96\%$ of $100$ synthetic data sets, and the false-failure rate
of a null at the threshold $Q$ is $0.03\%$ to $0.18\%$ in $4000$ synthetic
data sets, against a nominal $0.1\%$.

\emph{Design.} Job \texttt{daop43opqrnc739aancg} on \textit{ibm\_fez} was
submitted on 2026-09-21 at 20:22:36~UTC. The chain is 141, 142, 143, 144, 145,
146, 147, 148, 149, 150, 151, 138, 131, 130, 129, 118, 109, 110, 111, 98, the
qubits of the first job in reverse order, with the germ at chain position
$x_0=17$, physical qubit $110$. The reported readout errors on the rail
region were $0.29\%$ to $0.85\%$ and the $T_1$ times $82$ to $198\,\mu$s. The
record length is $S=4$, the late twirl enumerates the $256$ Pauli
assignments with $128$ shots each, the background is measured, and the
circuits are interleaved in eight blocks, which gives $32\,768$ shots per
configuration and germ sign. Gate A is the Haar gate of the first job (seed
$17$ of \texttt{lightcone\_records.haar\_unitary}). Gate B (seed $97$) was
selected before any data were taken, as the gate with the largest noiseless
memory at both $d=1$ and $d=2$ among seeds $1$ to $119$ with dual-unitarity
defect above $0.5$, namely $0.120$ and $0.165$ against $0.052$ and $0.031$
for gate A. The circuits were emitted, transpiled and checked by the code of
the first job. The column labelled largest in Table~\ref{tab:memory} refers,
as registered, to the component that is largest in the noiseless prediction.

\emph{Noise.} A single model describes the chain, with a depolarizing
strength $\lambda$ and relaxation $\gamma_{\rm post}$ on the wires of the
strip after every layer, relaxation $\gamma_{\rm pre}$ after every readout,
and assignment errors $q_{01}$ and $q_{10}$, fitted jointly to the full
distributions of the three $d=0$ records. The fit gives $\lambda=0.054$,
$\gamma_{\rm post}=0.007$, $\gamma_{\rm pre}=0.019$, $q_{01}=0.0082$ and
$q_{10}=0.0037$, at $\chi^2/\mathrm{dof}=2.72$. These five numbers are then
inserted into the co-moving law at $d=1$ and $d=2$. The model is not perfect,
as its own $\chi^2$ shows, and the total $\chi^2=60.5$ on the $36$
components of Table~\ref{tab:memreg} reflects that. Every registered
threshold is nevertheless met.

\emph{Outside the registered criteria.} The witness $W_h$ of the
dual-unitary gate at $d=1$, on the new rail qubits, is
$(0.010,0.004,-0.005,0.006)$ with a standard error of $0.004$, null to within
$2.5\sigma$, and at $d=2$ it is null to within $1.3\sigma$. This was not
among the criteria and is reported as exploratory.

\emph{Repetitions.} The registration provided for repetitions on a second
chain or a second processor, each analysed on its own against the same
criteria. The same nine configurations, with the same gates, seeds, shot
numbers and interleaving, were submitted to \textit{ibm\_marrakesh} at
20:33~UTC (\texttt{daop93opqrnc739ab0og}) and to \textit{ibm\_kingston} at
20:37~UTC (\texttt{daopb1n8gn2s739o77g0}). Only the chains differ. On
\textit{ibm\_marrakesh} the chain is 33, 34, 35, 19, 15, 14, 13, 12, 11, 10,
9, 8, 7, 6, 5, 4, 3, 2, 1, 0, with the germ on qubit $2$, and on
\textit{ibm\_kingston} it is 93, 79, 73, 72, 71, 70, 69, 68, 67, 57, 47, 46,
45, 44, 43, 42, 41, 36, 21, 20, with the germ on qubit $36$. Both jobs
returned all $589\,824$ shots, and the registered script analysed them
without change (Table~\ref{tab:repeat}). The noise model fitted to the edge
records gives $\lambda=0.024$, $\gamma_{\rm post}=0.002$,
$\gamma_{\rm pre}=0.015$, $q_{01}=0.0036$ and $q_{10}=0.0055$ on
\textit{ibm\_kingston}, and $\lambda=0.041$, $\gamma_{\rm post}=0.011$,
$\gamma_{\rm pre}=0.060$, $q_{01}=0.0060$ and $q_{10}=0.0045$ on
\textit{ibm\_marrakesh}. The calibration records in force when the two
repetitions ran, last updated at 01:28~UTC on \textit{ibm\_marrakesh} and
at 20:15~UTC on \textit{ibm\_kingston}, agree with the rail values stored
at submission (Table~\ref{tab:calsum}). A rerun of the registered analysis
on the stored records of the three jobs reproduces every number of their
reports to the precision quoted here.

\begin{table}[tbh]\footnotesize
\caption{The registered memory demonstration on \textit{ibm\_fez} and its
repetitions on \textit{ibm\_marrakesh} and \textit{ibm\_kingston}, each
analysed on its own by the registered script against the registered
criteria. Displaced Haar records give $\chi^2_0\,/\,\chi^2_{\rm law}$ as in
Table~\ref{tab:memory}. The nulls are the five records predicted to carry no
germ-odd memory. Significances refer to $\Delta(3,+)$ for gate A and to the
component largest in the noiseless law for gate B. The noise model row gives
$\chi^2/\mathrm{dof}$ of its fit to the edge records. The Markov rows give
$\chi^2/\mathrm{dof}$ of the step-dependent two-state chain, the largest
value among the three edge records and the dual-unitary record at $d=1$, and
the values for both Haar gates at $d=1$.}
\label{tab:repeat}
\begin{ruledtabular}
\setlength{\tabcolsep}{3pt}
\begin{tabular}{lccc}
 & \textit{ibm\_fez} & \textit{ibm\_marrakesh} & \textit{ibm\_kingston}\\ \hline
germ qubit & 110 & 2 & 36\\
noise model & 2.72 & 10.49 & 1.94\\
nulls, largest $\chi^2$ & 12.4 & 3.7 & 6.6\\
Haar A, $d=1$ & 71.6\,/\,7.3 & 94.7\,/\,11.0 & 46.9\,/\,12.1\\
Haar A, $d=2$ & 34.0\,/\,9.5 & 63.2\,/\,20.9 & 30.2\,/\,0.6\\
Haar B, $d=1$ & 251.8\,/\,5.9 & 623.1\,/\,115.8 & 231.3\,/\,21.1\\
Haar B, $d=2$ & 356.5\,/\,14.3 & 273.8\,/\,48.5 & 606.8\,/\,8.0\\
gate A, $d=1$ & $7.3\sigma$ & $9.6\sigma$ & $5.1\sigma$\\
gate B, $d=1$ & $9.8\sigma$ & $15.7\sigma$ & $11.3\sigma$\\
gate B, $d=2$ & $17.6\sigma$ & $14.6\sigma$ & $23.4\sigma$\\
Markov, nulls & 1.39 & 1.00 & 1.86\\
Markov, Haar A, B & 11.6, 30.3 & 17.3, 53.2 & 10.0, 24.9\\
criteria met & all & K1, K2, K5 & all but K3\\
\end{tabular}
\end{ruledtabular}
\end{table}

\section{Markov-structure fits}
\label{app:markov}

The model is Eq.~\eqref{eq:chain} with per-bit readout assignment errors
$q_{01}$ and $q_{10}$ common to all steps. Its parameters are the first-step
mean $\mu_1$, one per germ sign, the kernel $(t_h,s_h)$, either common to all
steps or one per step, and the two assignment errors. They are fitted by
maximizing the multinomial likelihood of the full distribution of $2^S$
outcomes under the constraint $|t_h|+|s_h|\le1$, from several starting
points. Goodness of fit is measured by Pearson's $\chi^2$ over cells with
expected count above $5$, with one constraint per germ sign and one degree of
freedom per fitted parameter subtracted. Pooled values are
$\sum\chi^2/\sum\mathrm{dof}$.

The rails of Table~\ref{tab:rails} are the datasets R1 = r12\_clean, R2 =
r12\_v2, R3 = results\_n12, R4 = r20\_clean, R5 = r20\_generic, R6 =
inj\_0\_0, R7 = r20\_haar2, R8 = results\_n20, R9 = r50\_clean, R10 =
results\_n50, R11 = r\_big and R12 = r\_100. Records not pooled there behave
as follows. Records with a planted single-qubit rotation give $0.62$, $0.65$
and $2.83$ against $1.93$ for the control on the same chain, and the
interleaved paired runs give $2.64$ and $3.90$ against $1.64$ and $3.82$ for
the controls measured in the same jobs, so the plant leaves the quality of
the fit unchanged. At $4\times10^4$ shots the residual of planted and control
records alike reflects the single readout error that the model assumes for
all steps. One planted record on a chain without a same-chain control gives
$0.27$. Chains containing a qubit with a stuck readout fail even the
per-step test, at $110$ to $201$. Rails with unital noise injected off the
ray at $p>0$ give $0.84$ to $2.43$.

\end{document}